\documentclass{aa}
\usepackage{graphicx}
\usepackage{caption}
\usepackage{subcaption}
\usepackage{placeins}
\usepackage{txfonts}
\usepackage{xcolor}
\usepackage[breaklinks,colorlinks,urlcolor=blue,citecolor=blue,linkcolor=blue]{hyperref}
\usepackage{booktabs}
\usepackage{multirow}
\usepackage{hyperref}

\begin{document}

   \title{Extreme ultraviolet late-phase flares as observed by EVE and AIA on board the Solar Dynamics Observatory}

   \author{S. Ornig\inst{1} \fnmsep \inst{2}\and
          A. M. Veronig\inst{3} \fnmsep \inst{4}\and K. Dissauer \inst{5}}

   \institute{Institute of Theoretical Astrophysics, University of Oslo,
              PO Box 1029 Blindern, 0315 Oslo, Norway\\
              \email{sascha.ornig@astro.uio.no}
         \and
             Rosseland Centre for Solar Physics, University of Oslo,
              PO Box 1029 Blindern, 0315 Oslo, Norway
         \and
             University of Graz, Institute of Physics, Universitätsplatz 5, 8010 Graz, Austria
         \and
             University of Graz, Kanzelhöhe Observatory for Solar and Environmental Research, Kanzelhöhe 19, 9521 Treffen, Austria
         \and
             NorthWest Research Associates, 3380 Mitchell Lane, Boulder 80301, CO, USA
             }

   \date{Received 03 July 2025 / Accepted 19 August 2025}

  \abstract
   {Extreme ultraviolet (EUV) late-phase (ELP) flares exhibit a second peak in warm coronal emissions minutes to hours after the main peak of the flare. This phase is all but negligible, and it is still poorly understood what role it plays across the solar cycle and what governs it.}
   {We present a statistical analysis of ELP flares over four years between May 2010 and May 2014 based on properties such as eruptivity, magnetic configuration, and late-phase duration, delay, and strength in order to understand what influences the likelihood of this class of flares and their behavior on a general scale.}
   {We primarily made use of data from the Solar Dynamics Observatory (SDO) Extreme ultraviolet Variability Experiment (EVE), as well as complementary spatial information provided by the Atmospheric Imaging Assembly (AIA), to assess relationships between the various parameters and to see if ELP flares differ from the general flare population. We quantified the criteria for ELP flare definition and determined the characteristics of the flares.}
   {Our analysis shows that about 10\% of all flares with a GOES class $\geq$C3.0 experience an EUV late phase (179 out of 1803). This percentage decreases from solar minimum to solar maximum. C-class flares are considerably less likely to be identified as ELP flares than their higher-energy counterparts, which is in line with previous investigations. The majority of this type of flare are confined (67\%), more so than in the general flare population ($\geq$C5.0). There appears to be a (linear) relationship between the late-phase delay and its duration. The ratio of the emission peak of the late and main flare phase lies between 0.3 and 5.9, and exceeds 1 in 71.5\% of cases, which is considerably higher than previously reported.}
   {}

    \keywords{Sun --
                solar flares -- extreme-ultraviolet -- late-phase
               }

    \titlerunning{EUV late-phase flares observed by SDO}
    \maketitle


\section{Introduction}\label{sec:intro}


Flares are localized sudden brightenings observable throughout the entire electromagnetic spectrum, lasting from minutes to hours. They can accelerate particles to high energies and heat plasma, and are often accompanied by coronal mass ejections \citep[e.g., reviews by][]{2011SSRv..159...19F,2011LRSP....8....6S,2017LRSP...14....2B}. The CSHKP~model (or Standard Flare Model), named after \cite{1964NASSP..50..451C}, \cite{1966Natur.211..695S}, \mbox{\cite{1974SoPh...34..323H}}, \cite{1976SoPh...50...85K}, was developed to explain these observations. In this model, a large magnetic loop rises from below the photosphere to coronal heights and becomes pinched at its base. Discussions of the 3D extension of this model can be found in \cite{2011IAUS..273..233A} and \cite{2017JPlPh..83a5301J}. Magnetic reconnection sets in, restructuring the magnetic field and releasing energy. Accelerated particles (electrons and ions) travel along the magnetic loop to the chromosphere, where the high density causes the particles to lose energy through Coulomb collisions, thus heating the surrounding plasma. Due to the fast increase in plasma pressure, the heated plasma impulsively expands along the loop toward the corona, a process known as chromospheric evaporation \citep[e.g., ][]{1985ApJ...289..414F}. This process causes the enhanced flare radiation in the soft X-ray (SXR) and extreme ultraviolet (EUV) parts of the electromagnetic spectrum \mbox{\citep{2017LRSP...14....2B}}. The cooling, plasma-filled loops are known as post-flare loops.\\
\indent The Extreme ultraviolet Variability Experiment \citep[EVE;][]{2012SoPh..275..115W} on board the Solar Dynamics Observatory \citep[SDO;][]{2012SoPh..275....3P} records full-disk solar irradiance measurements in the EUV. When studying the Fe~XVI~33.5~nm emissions (corresponding to a temperature of close to 3~MK; see Table~\ref{tab:lines}) observed by EVE, \cite{2011ApJ...739...59W} found that certain flares exhibit a second EUV peak minutes to hours after the main peak. Furthermore, they found that no corresponding \mbox{X-ray} emissions were detected, which would indicate another episode of impulsive energy release. This distinction from other flare phases led them to term this phenomenon the EUV late phase (ELP). The late-phase peak can sometimes even have a larger amplitude than the main peak. This property was labeled an extremely large EUV late phase, or extremely large ELP for short, by \cite{2015ApJ...802...35L}.\\
\indent Originally, \cite{2011ApJ...739...59W} defined four criteria for ELP flares: First, some minutes to a few hours after the GOES SXR peak, the warm coronal emissions (especially Fe~XV and Fe~XVI at about 3~MK) show a second peak. Second, there should be no substantial enhancements of the hot coronal (e.g., Fe~XX) and GOES \mbox{X-ray} emissions during this peak. Third, an eruption should be associated with the respective event, observed in SDO/AIA images, as a white-light CME in coronagraphs, or as a coronal dimming in the Fe~IX~17.1~nm emission line. Fourth, a second system of loops should be visible in AIA images, which are required to show some spatial and temporal separation from the post-flare loops.\\
\indent In more recent years, the third criterion (concerning eruptivity) has been mostly neglected, since confined flares have also been found to sometimes exhibit an ELP \mbox{\citep[e.g.,][]{2015ApJ...802...35L,2020ApJ...890..158C,2018ApJ...863..124D}}.\\
\indent There are currently two pathways of explanations for the appearance of the EUV late phase in the literature \mbox{\citep[e.g.,][]{2020ApJ...890..158C,2018ApJ...857...99D}}. One is that since the ELP flare stems from larger loops than the gradual phase, the cooling time of those loops is longer, resulting in a temporal delay \citep[e.g.,][]{2013ApJ...768..150L,2013ApJ...778..139S,2017A&A...604A..76M}. The second explanation is that some additional heating mechanism \citep[examples of which include a failed eruption or an additional phase of magnetic reconnection involving somehow weaker magnetic fields; see][]{2011ApJ...739...59W,2012arXiv1202.4819H,2013ApJ...773L..21D,2013ApJ...778..139S,2019ApJ...878...46Z,2022ApJ...932...53Z} provides an extra energy input into the region, resulting in heating of the plasma (but not to hot flaring temperatures) and creating the late phase. In general, both mechanisms can play a role in a single flare as well, meaning they are not mutually exclusive \citep[e.g.,][]{2015ApJ...802...35L,2021ApJ...916...37Z}.\\
\indent The first statistical analysis of ELP flares was performed by \cite{2011ApJ...739...59W}. They analyzed 191~flares between May~2010 and March~2011, which they chose based on the visibility of their post-flare loops. They found that 25 of the flares (about 13\%) produced an EUV late phase. For \mbox{C-class} flares, the percentage was 9.5\% (16 of 169), for \mbox{M-class} flares 42.9\% (9 of 21), and for \mbox{X-class} flares 0\% (0 of 1). In a follow-up study, \cite{2014SoPh..289.3391W} compared 3~years of EVE observations with the GOES \mbox{X-ray} light curves of some ELP flares to estimate how the ELP flare probability changes. Using dual-decay \mbox{X-ray} flares as proxies for ELP flares (between 1974 and 2013), they found that ELP flares appear more frequently prior to and after the solar minimum, but occur significantly less frequently during the cycle maximum.\\
\indent \cite{2020ApJ...890..158C} investigated 55~M- and \mbox{X-class} flares between May~2010 and May~2014. They chose flares with a substantial EUV late phase in the EVE profiles and a location within 45$^\circ$ of the disk center. Among other factors, they looked at the eruptivity and flare-ribbon morphology. They found that 19~events produced circular ribbons, 23 had the classical two-ribbon morphology, and 13 had a complex ribbon configuration. Regarding eruptivity, 22~flares in their analysis were eruptive, with complex-ribbon flares showing a larger fraction of eruptive flares than circular- and two-ribbon flares. On the other hand, they found that two-ribbon flares are more likely to produce extreme ELP events (i.e., ELP flares that show a ratio of late-phase peak to main-phase peak $>$1) than the other morphologies (48\% vs. 37\% and 31\% for circular- and complex-ribbon flares, respectively).\\
\indent The aims of the present study are to (i) increase the statistical significance by analyzing a larger set of flares, (ii) quantify the late-phase criteria, (iii) investigate the CME-association rate among ELP flares, and (iv) identify if (magnetic) active region (AR) configurations exist that are more likely to exhibit ELP activity.


\section{Instruments and data} \label{sec:instruments_and_data}


\subsection{Instruments} \label{subsec:instruments}

The main instruments used for this study are EVE and the Atmospheric Imaging Assembly \citep[AIA;][]{2012SoPh..275...17L} on board SDO. EVE takes irradiance measurements of the full solar disk at a high cadence (10~s) and good spectral resolution of 0.1~nm \citep{2012SoPh..275..115W}. The primary measurements by EVE are made via the Multiple EUV Grating Spectrographs (MEGS), which consist of \mbox{MEGS-A} and \mbox{MEGS-B}. The former is used for the spectral range between 5~and 37~nm, while the latter is designed for measurements in the 35~to 105~nm range \mbox{\citep{2004SPIE.5563..182C}}. \mbox{MEGS-A} suffered a CCD camera power anomaly, so its observations are limited to the time between May~1,~2010 and May~26,~2014.\\
\indent The lack of spatial information in the EVE irradiance measurements can be partially overcome by combining them with EUV imagery from AIA. AIA provides full-disk images of the photosphere, chromosphere, transition region, and corona up to 1.5~$R_\odot$ at a 12~s time cadence at multiple wavelengths, most of which overlap with measurements by EVE \citep{2015ApJ...802...35L}. The spatial resolution AIA provides is 1.5~arcsec.


\subsection{Data products and preparation} \label{subsec:data_prods_and_prep}

\begin{figure}
    \centering
    \includegraphics[width=0.50\textwidth]{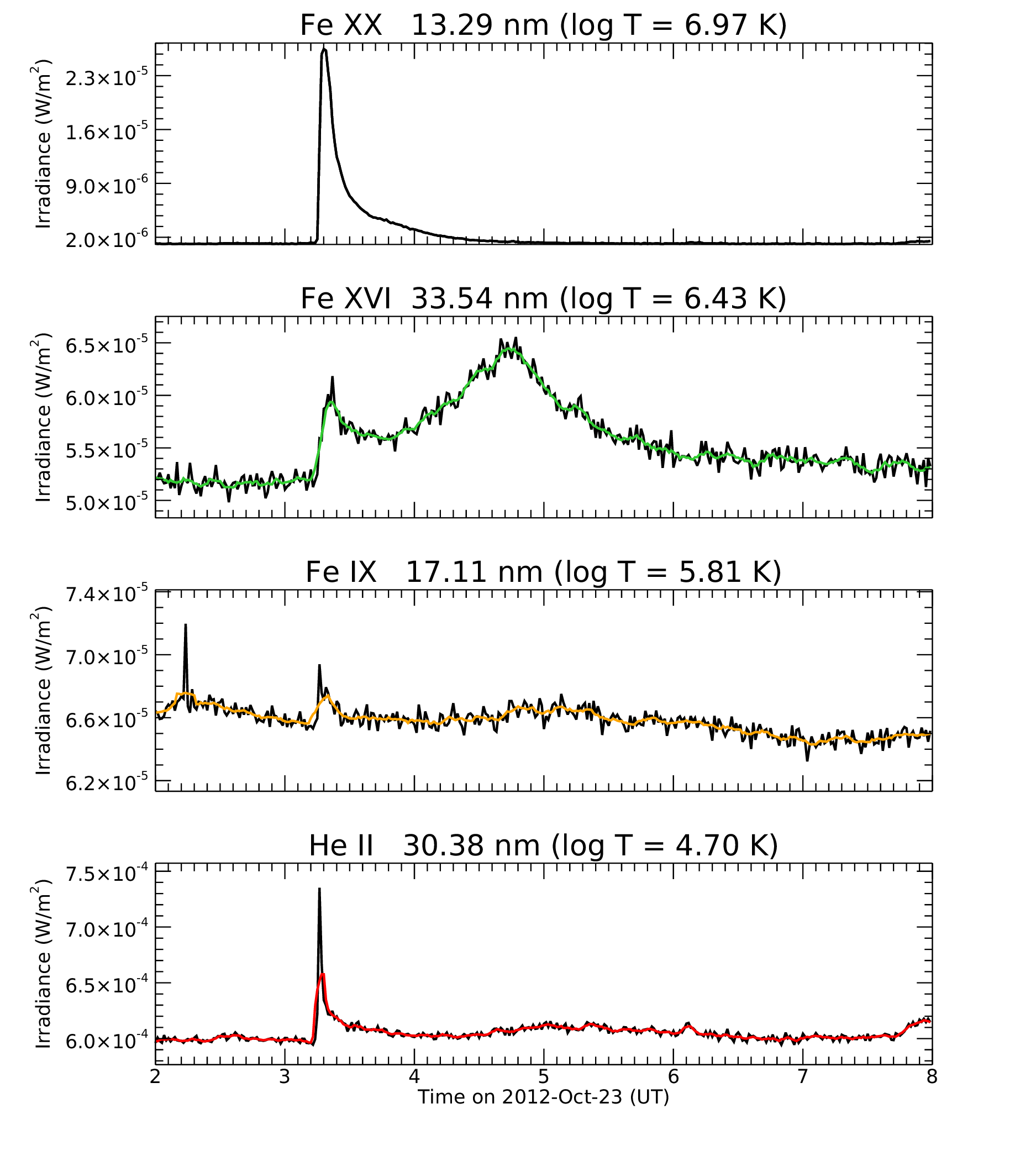}
    \caption{Unsmoothed (black) and smoothed (green, yellow, and red) light curves of the EVE emission lines from Table~\ref{tab:lines} for the X1.8~flare on October~23,~2012. The light curves are arranged according to the corresponding line formation temperature (given above each panel), decreasing from top to bottom. The Fe~XX light curves were not smoothed.}
    \label{fig:eve_raw}
\end{figure}

EVE~level~2~data products are available in two different forms: spectra (EVS) and spectral lines (EVL). The combined spectrum readings from the two spectrographs, \mbox{MEGS-A} and -B, make up the level~2~spectra. The irradiances are all adjusted to a normalized distance of 1~AU \citep{2012SoPh..275..115W}. Table~\ref{tab:lines} (upper half) shows the EVE emission lines relevant for the analysis in this paper. They were chosen in accordance with \cite{Woods_2011a}: (i) The Fe~XX~13.3~nm emissions (\mbox{$T \approx 10$}~MK) generally behave similar to the GOES emissions and are indicative of the gradual phase of the flare, (ii) the Fe~XVI~33.5~nm emissions (\mbox{$T \approx 3$}~MK) exhibit the late-phase phenomenon best, (iii) the Fe~IX~17.1~nm emissions (\mbox{$T \approx 0.6$}~MK) sometimes show a substantial decrease in irradiance (coronal dimming) which has been found to correlate well with CMEs \citep{2014ApJ...789...61M,2016ApJ...830...20M,2019ApJ...874..123D}, and (iv) the He~II~30.4~nm emissions (\mbox{$T \approx 60,000$}~K) indicate whether a strong impulsive phase is present in the flare \mbox{\citep{2011ApJ...739...59W}}.\\
\indent AIA~level~1~data were used for the spatial analysis. In accordance with EVE, we chose the 33.5~nm channel (peak formation temperature $T \approx 2.5\times10^6$~K), as it most closely covers the same spectral region as the EVE~33.5~nm channel. For each event, a subregion on the solar surface that covered the flare region was selected. The cadence was chosen to be 1~minute. Standard SDO/AIA routines ($aia\_prep$) were used for data reduction. The AIA~160~nm channel was utilized to identify the flare-ribbon morphology (Sect.~\ref{subsubsec:ribbons}).\\
\indent The EVE data were downloaded with the original 10-second cadence. All emission lines used in this paper (13.3~nm, 17.1~nm, 30.4~nm, and 33.5~nm) are included in the EVL product. To improve the signal-to-noise ratio, the 10-second data were time-integrated and re-binned to a 1-minute cadence. Furthermore, since emissions in lower-temperature channels tend to show more pronounced noise (e.g.,~Fig.~\ref{fig:eve_raw}, black curves), we applied central moving averages with varying box sizes. The 13.3~nm emissions show little noise, so no smoothing was applied. The 33.5~nm emissions exhibit a modest amount of noise, so \mbox{$n = 7$} (i.e.,~an average over 7~minutes) was chosen. For the 17.1~nm series, we used \mbox{$n = 9$}, and for the 30.4~nm~series \mbox{$n = 5$}. An example of smoothed light curves can be found in Fig.~\ref{fig:eve_raw} (colored lines). The results we present in this article are based on the EVE version~8 pre-beta data provided by the EVE team, rather than the (at the time) publicly available version~7,\footnote{Data access: \url{https://lasp.colorado.edu/eve/data_access/eve_data/products/level2/}.} due to large data gaps in the version~7 data.

\section{Methods}\label{sec:methods}


\subsection{ELP criteria and flare selection}\label{subsec:elp_crit}

\begin{figure}
    \centering
    \includegraphics[width=0.47\textwidth]{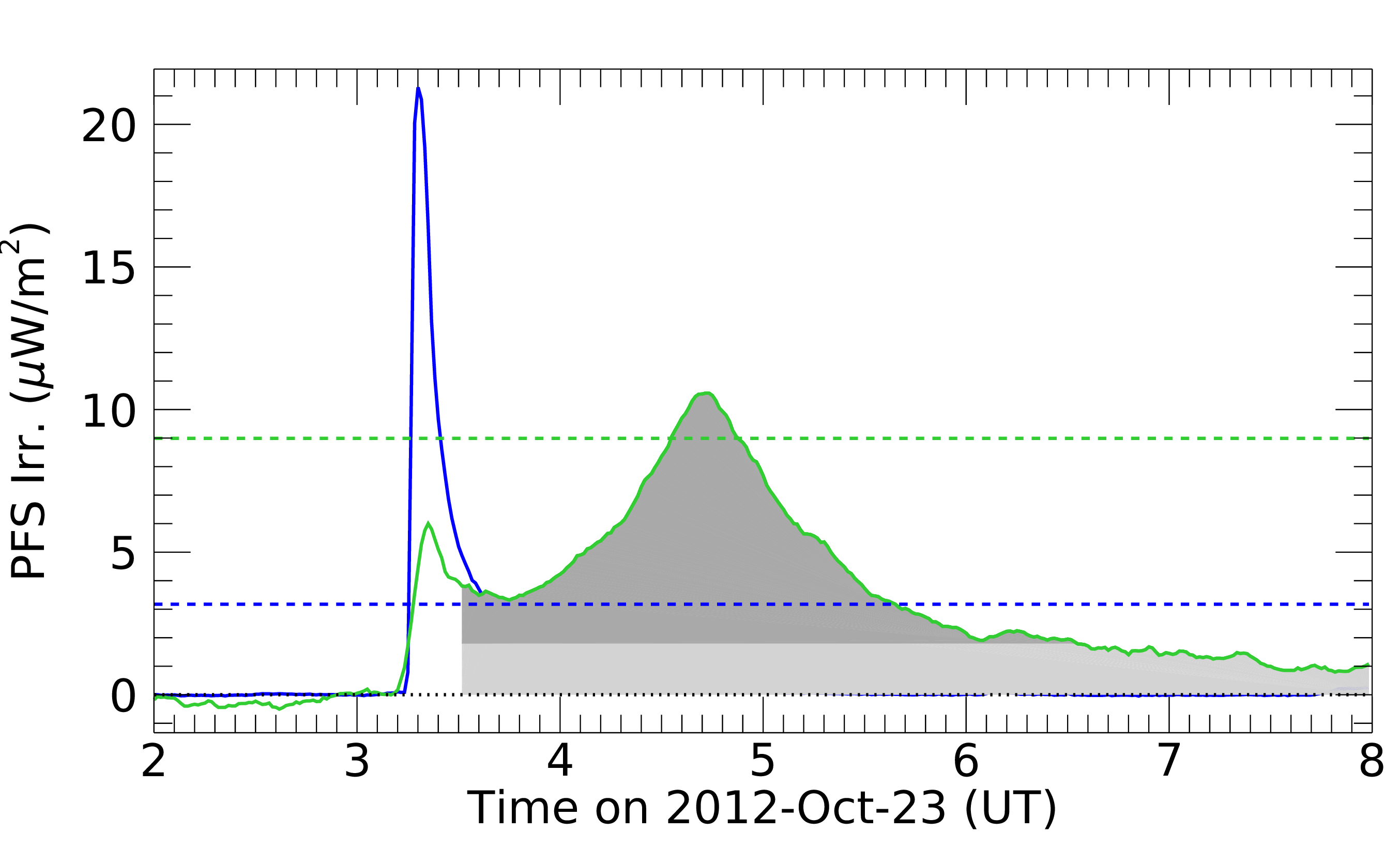}
    \caption{Visual representation of the ELP criteria applied in this study. Depicted are the PFS irradiances of the EVE Fe~XX~13.29~nm (blue) and Fe~XVI~33.54~nm (green) emission lines for the X1.8~flare on October~23,~2012. The horizontal dotted line represents the pre-flare level. The shaded gray areas correspond to the region in the Fe~XVI light curve that satisfies criterion~1 (darker shade of gray) and criterion~2 (lighter and darker shade of gray together) given in Sect.~\ref{subsec:elp_crit}. The horizontal dashed green line marks the limit corresponding to criterion~3, whereas the horizontal dashed blue line depicts the limit for criterion~4.}
    \label{fig:eve_crit}
\end{figure}
The criteria for the presence of an EUV late phase defined by \cite{2011ApJ...739...59W} are laid out in Sect.~\ref{sec:intro}. For this work, we revised and quantified those late-phase criteria as follows: (1) The pre-flare-subtracted (PFS) irradiance during the potential late-phase maximum should have at least 30\% of the value it showed during the main-phase maximum, (2) there should be at least 10~minutes between the main-phase and late-phase maxima (this value was chosen based on visual inspection of the light curves of a multitude of potential ELP flares), (3) there should be a local minimum between the two maxima and the PFS irradiance level at that local minimum should drop below 85\% of the late-phase maximum, (4) in analogy to \cite{2011ApJ...739...59W}, there should be no substantial enhancements of the hot coronal emissions (Fe~XX~13.3~nm) within 15~minutes before the late-phase peak, where substantial in this case means a local maximum in the Fe~XX~13.3~nm emission line with a PFS irradiance value greater than 30\% of the Fe~XVI~33.5~nm late-phase maximum, and (5) the emissions during the late phase should stem from the same AR as during the main phase of the flare.\\
\indent A visual representation of criteria~1--4 is shown in Fig.~\ref{fig:eve_crit}. The first criterion aims to establish a base for what is deemed a substantial second peak. Using the second criterion, we try to eliminate flares that show a temporary decrease in emissions, for example due to an erupting flux rope (which may be optically thick enough to block out emissions in higher-energy passbands), which could potentially result in a classical flare being falsely identified as an ELP flare. With the third criterion, we particularly intend to distinguish between events that show a roughly constant (or very slowly decreasing) level of emission over a longer period of time, and ELP events (which produce a noticeable secondary maximum). Criterion~4 is our take on quantifying substantial hot coronal emissions. Since the Fe~XX emissions during a flare usually exceed the Fe~XVI emissions in terms of PFS irradiance (or are at least similar in strength), a classical flare would not result in small Fe~XX enhancements creating large Fe~XVI emission peaks. Finally, the fifth criterion is important to eliminate emissions from different regions of the Sun being identified as the late phase in the full-Sun EVE data.\\
\indent As suggested by \cite{2013ApJ...768..150L} and further implemented by \cite{2015ApJ...802...35L}, \cite{2018ApJ...863..124D}, and \mbox{\cite{2020ApJ...890..158C}}, among others, we do not demand a priori that the potential ELP flare is associated with an eruption. Further, we do not require a second set of loops to be present, in order to avoid any a priori presumption on the physical reasons for the ELP occurrence.\\
\indent To compare our criteria with those previously reported in the literature, we initially applied them to flares that were identified as ELP flares by \mbox{\cite{2011ApJ...739...59W}}, \cite{2013ApJ...768..150L}, \cite{2018ApJ...863..124D}, and \cite{2020ApJ...890..158C} according to their criteria. \mbox{\cite{2011ApJ...739...59W}} studied flares of class C2.0 and above between May~2010 and March~2011. However, they only analyzed flares where the post-flare and ELP loops were well visible. Out of the 191~flares (169~C-, 21~M-, and 1~\mbox{X-class} flare) that qualified, 25 (16~C- and 9~\mbox{M-class} flares) were identified as ELP flares (13.1\%). \cite{2013ApJ...768..150L} provided info on two ELPs, one M2.9 and one M1.4~flare, while \cite{2018ApJ...863..124D} looked at an M1.2~ELP flare. Unfortunately, these two publications did not disclose the criteria they used for identification. The most extensive set of ELP flares was provided by \mbox{\cite{2020ApJ...890..158C}}, who analyzed all 473~M- and \mbox{X-class} flares between May~2010 and May~2014. In contrast to the ELP criteria of \cite{2011ApJ...739...59W}, they did not require the potential ELP flare (i) to be eruptive, or (ii) to display two separate loop systems. They found 55~flares exhibiting a late phase (11.6\%), which is a slightly lower fraction than it was the case for \mbox{\cite{2011ApJ...739...59W}}. However, a higher ELP frequency near solar minimum was already proposed by \mbox{\cite{2014SoPh..289.3391W}}. In total, the aforementioned publications identified 67~EUV late-phase flares, 48 of which still qualify as such according to our definition. Our criteria are therefore more strict than those in these works.\\
\indent Since the compilation of ELP flares found in the literature still results in a very limited number of ELPs, we extended our dataset to all 1803~flares above GOES class C3.0 between May~1,~2010, and May~26,~2014, included in the EVE flare catalog developed by \mbox{\cite{2012PhDT........57H}}.\footnote{Available at \url{lasp.colorado.edu/eve/data_access/eve-flare-catalog/index.html}.} The flare selection and ELP determination process can be summarized in two steps: 1) We visually checked the time series plots of a variety of lines available in the EVE flare catalog covering a temperature range from roughly $5\times10^4$~K to $10^7$~K \mbox{\citep{2012PhDT........57H}}. All flares that show a second peak in EVE~33.5~nm emissions in these time series plots without corresponding peaks in EVE~13.3~nm or GOES measurements were considered as potential ELP candidates and analyzed in Step 2. The others were classified as non-ELP flares. 2) For the identified ELP flare candidates, we downloaded and processed the EVE data, and applied the late-phase criteria. Those that did not fulfill all the criteria were moved to the non-ELP flare group. Out of the 1803~flares above C3.0, 179 were found to satisfy the criteria as ELP flares (9.9\%). They form the dataset for the further study of ELP flare characteristics.
\begin{figure}
    \centering
    \includegraphics[width=0.36\textwidth,trim={0cm 0cm 0cm 3cm},clip]{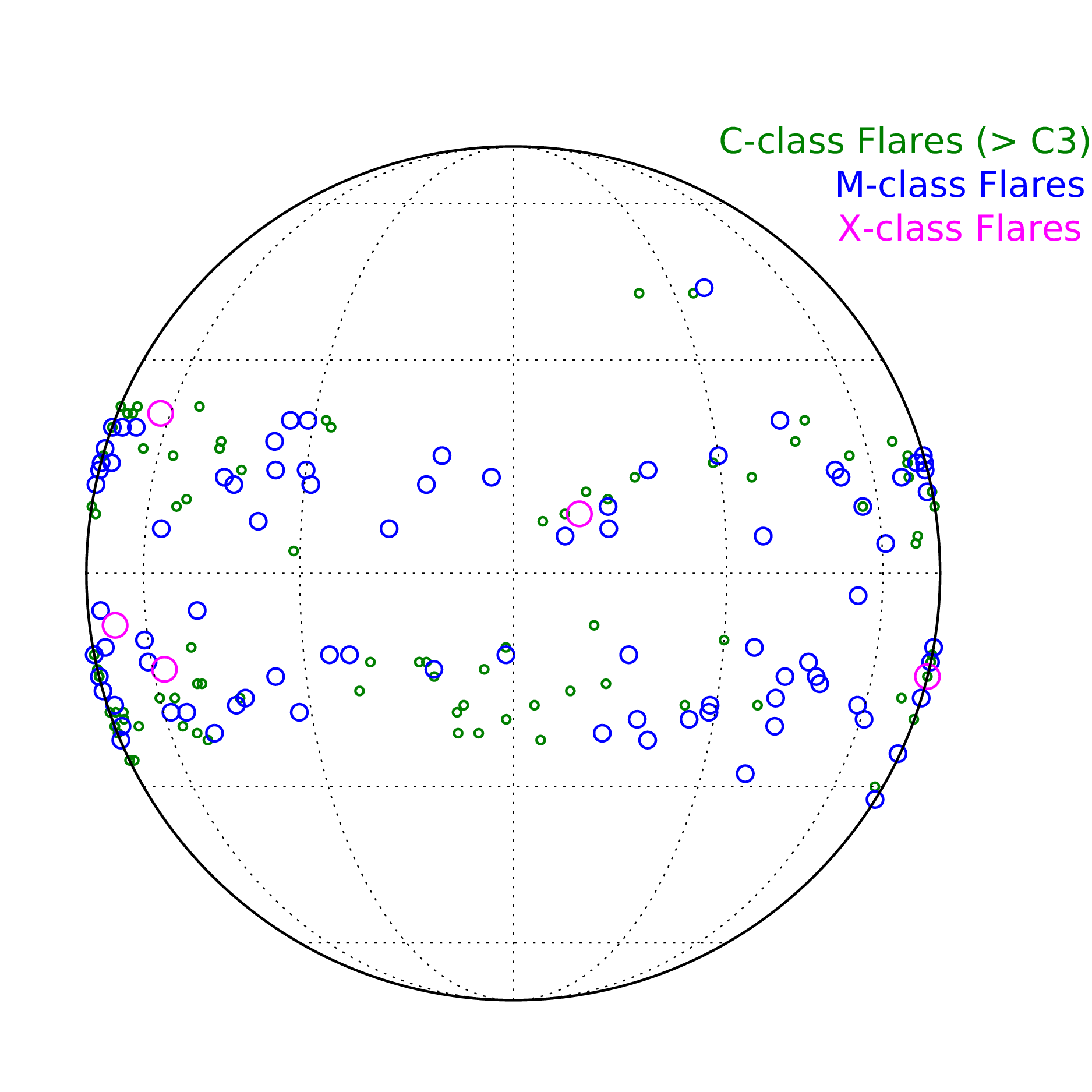}
    \caption{Visualization of the location of ELP flares on the solar surface, as viewed from a terrestrial perspective. \mbox{C-class} flares are marked with small green circles, \mbox{M-class} flares with medium-sized blue circles, and \mbox{X-class} flares with large magenta circles.}
    \label{fig:location}
\end{figure}


\subsection{Derivation of characteristic ELP properties}\label{subsec:props}

To analyze the flare set, we need to specify what parameters are of interest. To investigate whether there are obvious differences in ELP occurrence between the two solar hemispheres, we derive the flare location as described in Sect.~\ref{subsubsec:location}. The eruptivity (Sect.~\ref{subsubsec:erup}), flare-ribbon morphology (Sect.~\ref{subsubsec:ribbons}), and ELP productivity of ARs (Sect.~\ref{subsubsec:elp_prod}) help us identify different (magnetic) configurations associated with ELP flares, and if any patterns can be recognized in connection with them. Finally, we define the ratio of the late-phase to the main-phase peak (Sect.~\ref{subsubsec:peak_ratio}), as well as the delay (Sect.~\ref{subsubsec:delay}) and duration (Sect.~\ref{subsubsec:duration}) of the late phase to characterize each ELP flare.


\subsubsection{Location}\label{subsubsec:location}

For the determination of the heliographic location of flares, we again used the EVE flare catalog by \cite{2012PhDT........57H}. Therein, multiple statements about the location of a flare can be found. If available, we used the location according to the GOES flare catalog.\footnote{Can be queried within IDL or at \url{https://data.nas.nasa.gov/helio/portals/solarflares/webapp.html}.} If this information was not available, we took the location from the Heliophysics Event Knowledge (HEK) database.\footnote{Available at \url{www.lmsal.com/isolsearch}.} If no location was given in HEK, we used the location determined using the SAM instrument on EVE. A schematic representation of the location of all ELP flares studied in this work is shown in Fig.~\ref{fig:location}.


\subsubsection{Eruptivity}\label{subsubsec:erup}

The eruptivity of a flare can be defined in multiple ways. For this analysis, we focused on flares that produced a white-light CME observed in coronagraphs. We manually inspected the white-light images using the SOHO/LASCO~CME~catalog \citep{2004JGRA..109.7105Y}\footnote{Available at \url{cdaw.gsfc.nasa.gov}.} to determine whether a CME is associated with the ELP flare under consideration. Those that are accompanied by a CME are classified as eruptive flares, the other ones as confined flares.\\
\indent To infer the eruptivity of the general solar flare population and compare it with that of ELP flares, we used the flare catalog by \cite{2021ApJ...917L..29L},\footnote{Available at \url{dx.doi.org/doi:10.12149/101067}.} which records, among other parameters, the flare--CME connection for all flares $\geq$C5.0 between June~2010 and September~2017.


\subsubsection{Flare-ribbon morphology} \label{subsubsec:ribbons}

The emissions in the AIA~160~nm channel are representative of the upper photosphere, the chromosphere, and the transition region \citep{2012SoPh..275...17L}, where the footpoints of the flare loops are located. Since flare loops are not isolated but rather appear as flare arcades, the resulting signatures in the 160~nm channel are not single footpoints, but flare ribbons. According to the 2.5D~Standard~Flare~Model, we would expect two elongated, parallel flare ribbons observed during a classical flare. However, in reality they appear in very different and more complex shapes. In this paper, we follow the classification of \cite{2020ApJ...890..158C}, who divided flares based on their flare-ribbon morphology into the classical two-ribbon flares, as well as circular- and complex-ribbon flares.\\
\indent Two-ribbon flares show two more or less parallel, elongated ribbons that usually move apart over time. However, the angle of one ribbon with respect to the other can basically range anywhere from 0$^\circ$ (parallel) to almost 90$^\circ$ (perpendicular). An example is shown in Fig.~\ref{fig:ribbons} (top panel). Circular-ribbon flares feature a single ribbon that resembles a (half-) circle and is frequently open on one side. Figure~\ref{fig:ribbons} (middle panel) shows such a configuration. These ribbons may also have a dot-like footpoint in their centers or a hook-shaped ribbon connecting one of the ends of the open circle to a remote location \mbox{\citep{2012ApJ...760..101W,2020ApJ...890..158C}}. Finally, ribbons that fall outside both categories are referred to as complex (or irregular). This category also includes mixed flare-ribbon morphologies, which exhibit properties of both two-ribbon and circular-ribbon flares. Figure~\ref{fig:ribbons} (bottom panel) shows an example for a complex configuration.\\
\indent With regard to the flare ribbons and their shape and/or morphology, the position of a flare is also crucial. The ribbons become deformed by projection effects, which grow more pronounced the closer the flare is to the limb. At some point, the projection effects become so strong that it is impossible to determine flare-ribbon shapes. Therefore, we limited the investigation of flare-ribbon morphology to flares with a central meridian distance of $<$70$^\circ$.


\subsubsection{ELP productivity} \label{subsubsec:elp_prod}

The flare productivity of ARs may vary greatly. For instance, sunspot groups with a $\delta$- or \mbox{$\gamma$-configuration} are regarded to be more prone to creating large flares \citep{2000ApJ...540..583S,2019LRSP...16....3T}. Therefore, it seems reasonable to assume that ELP productivity may also vary between ARs of different complexity.\\
\indent In this work, we examine whether any of the possible configurations is more likely to produce ELP flares, rather than non-ELP flares. To accomplish this, we made use of the EVE flare catalog \citep{2012PhDT........57H}, which also includes AR associations including the Mount Wilson and McIntosh class for most of the flares included in the catalog. 319 of the 1803~flares did not have a class assigned to their respective sunspot regions, and are therefore left out of this part of the analysis.


\subsubsection{Main--to--late-phase peak ratio}\label{subsubsec:peak_ratio}

The peak ratio is defined as the PFS irradiance flux at the late-phase peak divided by the corresponding value at the main-phase peak (both in the EVE~33.5~nm irradiance measurements). Flares with a ratio $>$1, i.e., where the late phase has a larger peak value than the main phase, are termed extreme ELP flares \mbox{\citep{2015ApJ...802...35L}}.


\subsubsection{Late-phase time delay} \label{subsubsec:delay}

The delay between the main-phase maximum and the late-phase maximum was suggested to be a function of the loop length of the systems by \cite{2011ApJ...739...59W}. To determine the late-phase delay, we used the smoothed EVE~33.5~nm PFS irradiance and determined all local maxima. If multiple maxima occurred during the main and/or late phase, we investigated the corresponding AIA~33.5~nm images and light curves to determine which maxima correspond to the main- and late-phase peak. Finally, the time from the main maximum in the EVE~33.5~nm PFS irradiance to the late-phase maximum is then defined as the late-phase delay.


\subsubsection{Late-phase duration} \label{subsubsec:duration}

To investigate the duration of the late phase, we defined the minimum between the main-phase and late-phase maximum in EVE~33.5~nm measurements to be its starting point. The end is reached as soon as one of the two following criteria is met: (i) The PFS irradiance level of the Fe~XVI~33.5~nm emissions drops below 40\% of its late-phase peak, or (ii) a new flare of substantial strength, which visibly influences both the EVE~33.5~nm and AIA~33.5~nm measurements (some flares do not exhibit substantial emissions in this wavelength range), appears. In the latter case, the late-phase end is then defined as the start time of the new flare. ``Substantial strength'' here means that the peak GOES SXR flux of the new flare must be $>$10\% of the peak flux of the ELP flare and the new flare must be of class C1.0 or higher. AIA~33.5~nm measurements (images and subregion light curves) were used to determine the contribution of the new flare on the emission level (e.g., if a new flare with a strength classified as substantial by the measure given above happened, but AIA~33.5~nm measurements showed no major effect on the emissions, the end of the late phase was still determined by the 40\% criterion).


\subsubsection{Uncertainties}\label{subsubsec:uncertainties}

The uncertainties in the determined parameters are not uniform between events. Therefore, to analyze the uncertainties present in the late-phase delay and duration, we created AIA subregion light curves which were required to cover the whole region of the ELP flare as well as most of the AR that produced the flare under study. This is necessary to isolate the emissions coming from the flare region from those originating elsewhere on the Sun. Using the created light curve, the times of the main- and late-phase maximum, as well as the minimum in between and the end of the late-phase duration were calculated. The differences between these values and the EVE-calculated ones then help to estimate the uncertainties for both the late-phase delay and its duration. If there was no clear end to the late phase (e.g., due to another flare in the same region), the duration was estimated from the shape of the light curve before the onset of the new flare (the events for which this is the case are marked with a star in the flare list, see Sect.~\ref{sec:data_av} and Appendix~\ref{sec:flare_list}). An example is shown in Fig.~\ref{fig:eve_aia_uncertainty}: A new flare occurred prior to the decrease below the 40\%~threshold (see Sect.~\ref{subsec:elp_crit}) in AIA~33.5~nm emissions. The dashed blue line marks the estimated decrease in the AIA light curve if no new flare was occurring.\\
\indent To identify the uncertainties in the peak ratio, we varied the determined pre-flare level depending on the pre-flare activity, which usually shows fluctuations to some degree. The lower limit is defined as the determined pre-flare level minus $1\sigma$ (standard deviation of the unsmoothed irradiance within the pre-flare time window), while the upper limit is taken as the pre-flare level plus $1\sigma$. Calculating how the peak ratio changes as a function of pre-flare level establishes a good estimate for the resulting uncertainties in the parameters.

\section{Results} \label{sec:results}

\begin{figure}
    \centering
    \includegraphics[width=0.47\textwidth]{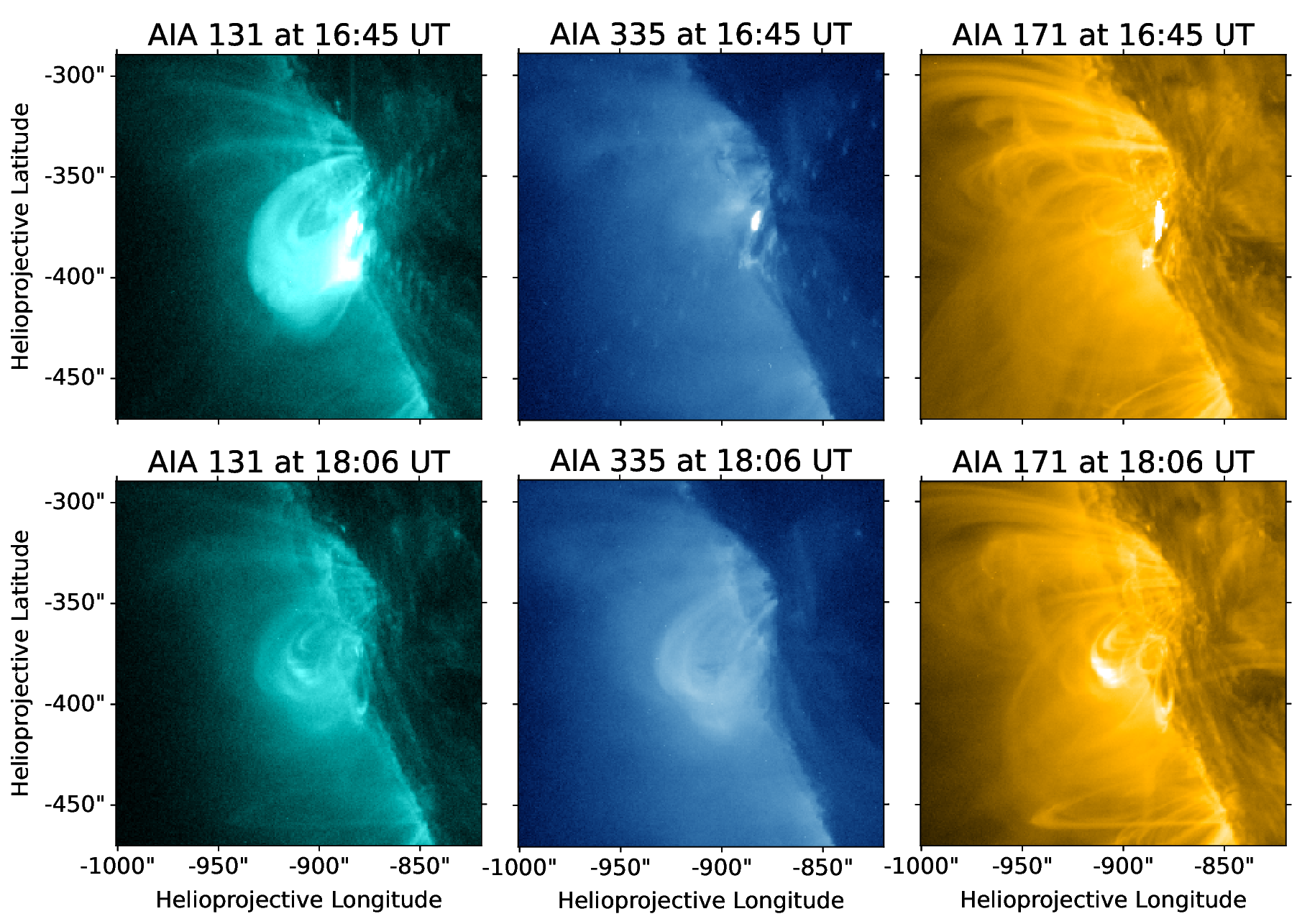}\\
    \includegraphics[width=0.47\textwidth]{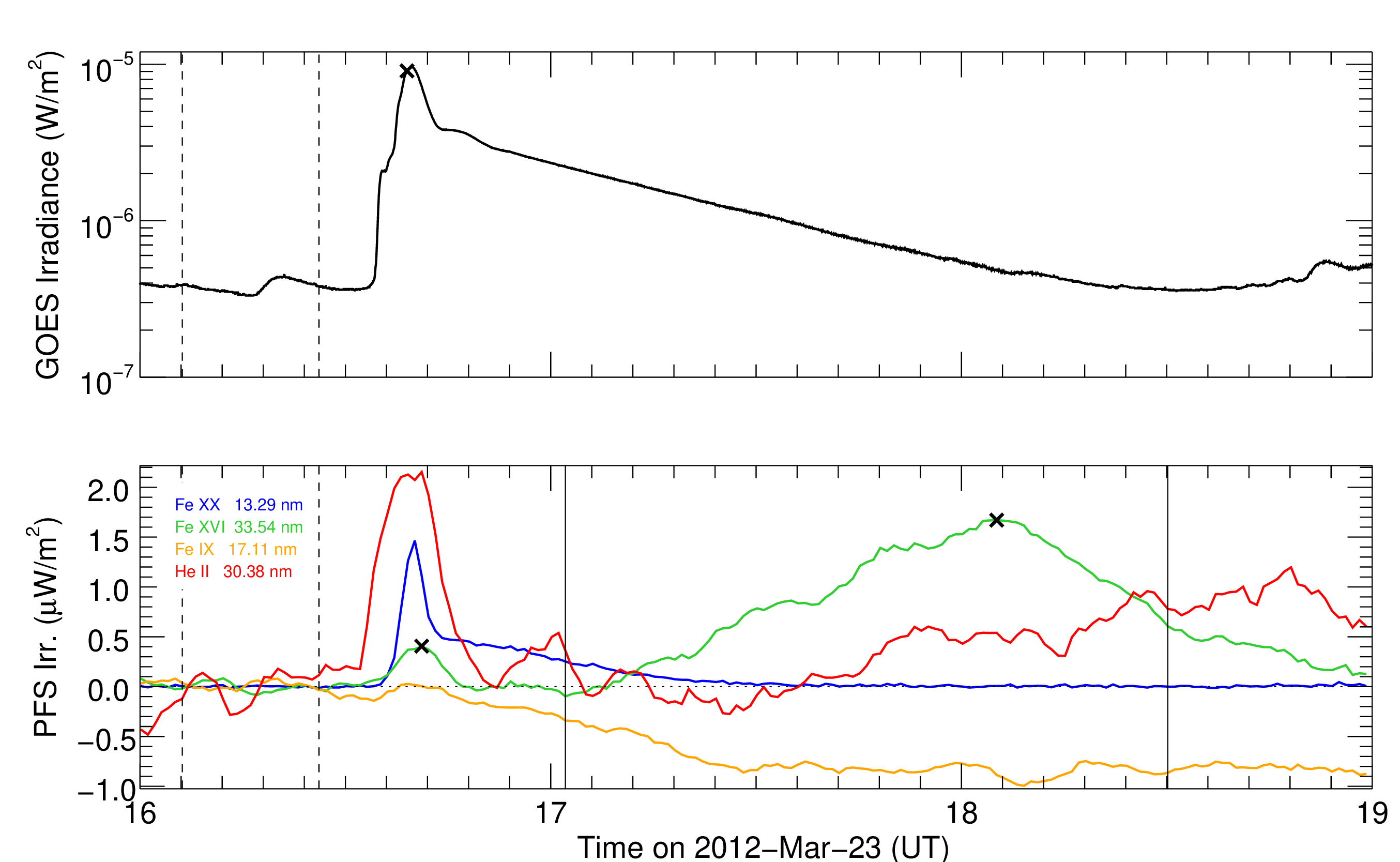}
    \caption{AIA images as well as GOES and EVE light curves for the confined C6.5 flare on March 23, 2012. Top two rows: Still images taken by AIA in the 13.1~nm (left column), 33.5~nm (middle column), and 17.1~nm (right column) filter. Middle panel: GOES 0.1--0.8~nm irradiance. The start and end of the pre-flare window used to calculate the PFS irradiances are shown by the two vertical dashed lines. Bottom panel: PFS irradiances of selected EVE emission lines: Fe~XX~13.29~nm (blue), Fe~XVI~33.54~nm (green), Fe~IX~17.11~nm (yellow), and He~II~30.4~nm (red). The PFS irradiance values for the He~II~30.4~nm line were divided by a factor of 3. The horizontal dotted line represents the zero level. The main-flare maximum and late-phase maximum are marked with a black cross. The vertical solid lines mark the beginning and end of the EUV late phase.}
    \label{fig:aia_eve_high_ratio}
\end{figure}

All flares identified as ELPs in this paper and their respective characteristics, which we analyze in more detail in the following sections, can be found in Sect.~\ref{sec:data_av} and Appendix~\ref{sec:flare_list}. Figure~\ref{fig:aia_eve_high_ratio} shows an example of an ELP flare, both from the perspective of AIA and as measured by GOES and EVE. As is often the case for ELP flares, after the peak in Fe~XX~13.3~nm emission, there is a steep decrease, followed by a more gradual decline. The Fe~XVI~33.5~nm emissions show a peak that is delayed by a few minutes compared to the Fe~XX peak. In this example, Fe~XVI shows a clear return to pre-flare levels, after which the late phase begins. The irradiance enhancements in Fe~XVI during the late-phase peak exceed those during the main flare phase. The upper panels of Fig.~\ref{fig:aia_eve_high_ratio} and the associated movie show that the late phase is caused by the cooling of hot plasma in loops that are substantially larger than the post-flare loops. Therefore, in this event the different cooling times may cause the delay between the main- and late-phase emissions.


\subsection{ELP-flare occurrence and eruptivity} \label{subsec:elp_prob_results}


\subsubsection{Probability of ELP-flare occurrence and evolution over the solar cycle}

\begin{figure}
    \centering
    \includegraphics[width=0.46\textwidth]{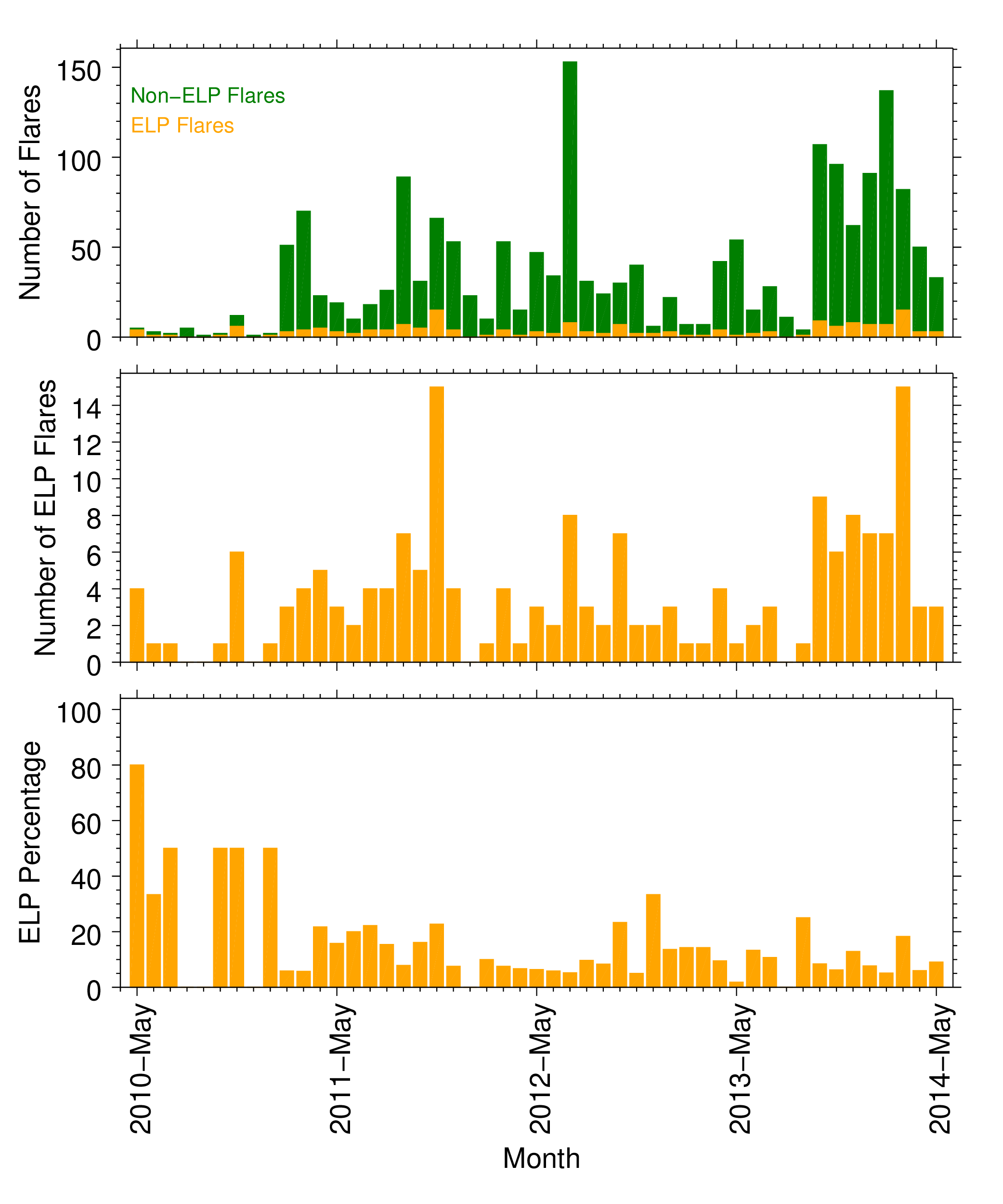}
    \caption{Monthly number of flares from May~2010 to May~2014. Top panel: Non-ELP flares $\geq$C3.0 (green) and ELP flares (orange). Middle panel: ELP flares. Bottom panel: Percentage of ELP flares.}
    \label{fig:flares_per_month}
\end{figure}

After applying the criteria laid out in Sect.~\ref{subsec:elp_crit}, 179 out of the total set of 1803~flares $\geq$C3.0 ended up qualifying as ELP flares (9.9\%). The breakdown according to GOES SXR classes is shown in Table~\ref{tab:elp_classes}. The absolute share of ELP flares from C- and M-class flares is comparable, at 89 and 85, respectively, while X-class flares make up 5 ELPs. On the other hand, the relative number of ELPs is much higher for M- and X-class flares (19.3\% and 16.1\%, respectively) as compared to the lower-energy C-class flares (6.7\%). As Fig.~\ref{fig:location} shows, there does not appear to be a preferred location for ELP-flare occurrence.\\
\indent It is worth noting that almost half of all ELP flares (83 of 179; 46.4\%) occurred during just 10 months as a result of two periods of high flaring activity from September 2011 to December~2011 (31 of 179; 17.3\%) and from October 2013 to March 2014 (52 of 179; 29.1\%). We further discuss the distribution of ELP flares across GOES SXR classes in Sect.~\ref{subsec:erup_results}.\\
\indent The monthly number of flares $\geq$C3.0 is shown with green bars in the top panel of Fig.~\ref{fig:flares_per_month} together with the number of ELP flares (orange bars).\footnote{The value for May 2014 is lower than the actual number of flares during this month, since the period of time we investigate is only up to May 26, 2014.} One can see that the ELP-flare frequency follows the overall flare frequency. We applied bootstrapping using 10,000~N-out-of-N random data pairs with replacement to calculate the Pearson correlation coefficient, $P_c$, and its confidence interval \citep{Efron_1994,Davison_1997}. We obtain \mbox{$P_c = 0.68$} and \mbox{$CI_{95\%} = [0.54,0.78]$} (95\%-confidence interval), which signifies a modest to strong positive linear relationship.\\
\newcommand{\specialcell}[2][c]{%
\begin{tabular}[#1]{@{}c@{}}#2\end{tabular}}
\begin{table}
    \caption{Number and percentage of ELP flares ($n_{ELP}$) and non-ELP flares ($n_{noELP}$) per GOES class between May 1, 2010, and May 26, 2014.}\label{tab:elp_classes}
    \centering
    \begin{tabular}{ |c|c|c|c|c|c| }
    \hline
    \specialcell{GOES\\Class} & $\mathrm{n_{tot}}$ & $\mathrm{n_{ELP}}$ & \specialcell{$\frac{\mathrm{n_{ELP}}}{\mathrm{n_{tot}}}$ (\%)} & $\mathrm{n_{noELP}}$ & \specialcell{$\frac{\mathrm{n_{noELP}}}{\mathrm{n_{tot}}}$ (\%)} \\
    \hline
    C & 1331 & 89 & 6.7 & 1242 & 93.3 \\
    M & 441 & 85 & 19.3 & 356 & 80.7 \\
    X & 31 & 5 & 16.1 & 26 & 83.9 \\
    \hline
    Sum & 1803 & 179 & 9.9 & 1624 & 90.1 \\
    \hline
    \end{tabular}
    \tablefoot{Only flares $\geq$C3.0 are taken into account.}
\end{table}
\begin{figure}
    \centering
    \includegraphics[width=0.49\textwidth]{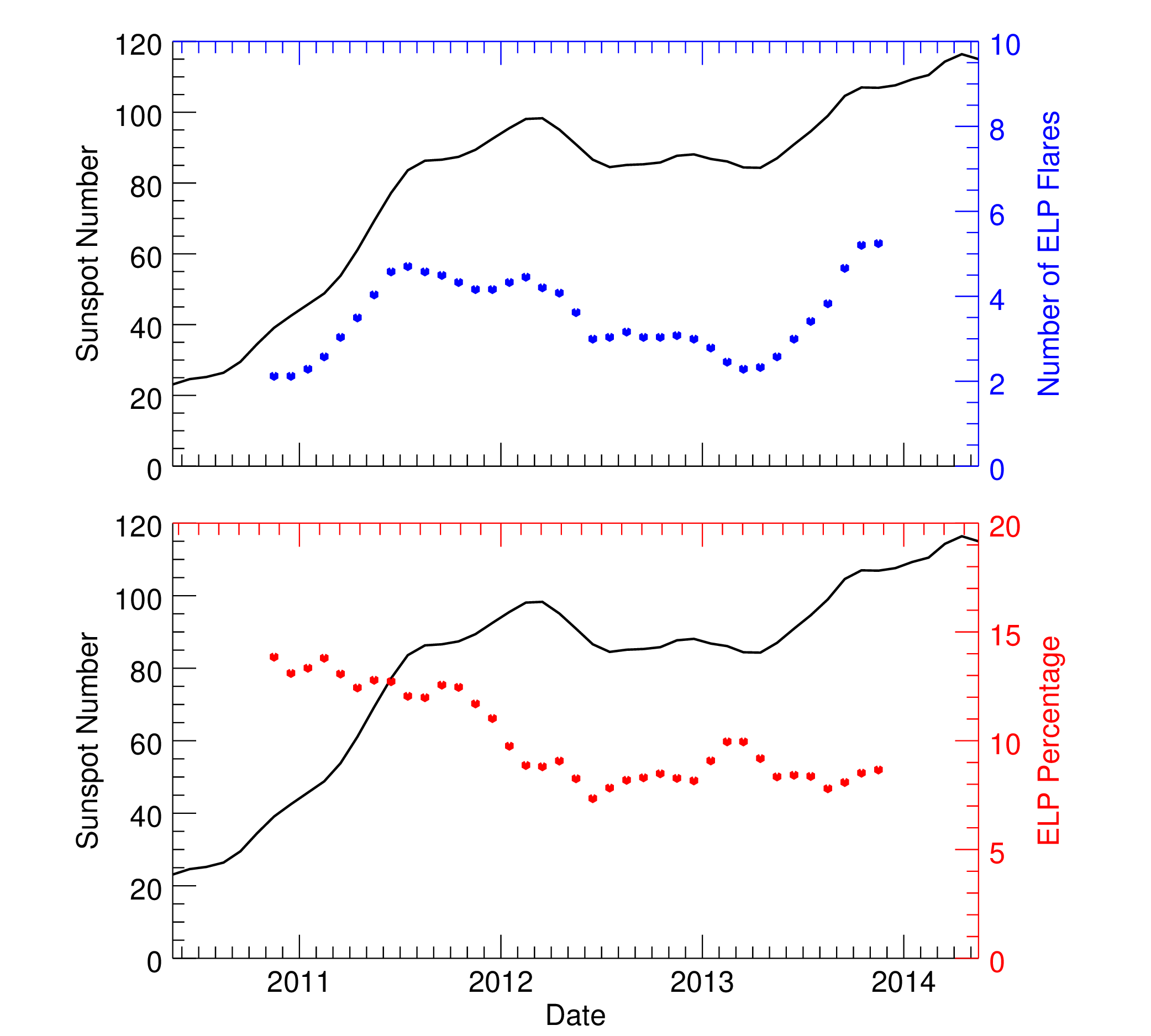}
    \caption{13-month smoothed monthly total number (top) and percentage (bottom) of ELP flares relative to the overall number of flares $\geq$C3.0 compared to the 13-month smoothed monthly mean total sunspot number (black curve) as a function of time between May 2010 and May 2014.}
    \label{fig:elp_vs_cycle}
\end{figure}
\indent The total number of ELP flares per month is separately depicted in the middle panel of Fig.~\ref{fig:flares_per_month}. Complementary information is provided by the bottom panel, which shows the percentage of ELP flares. During the first nine months (May~2010 to January~2011), i.e., close to the solar minimum, the portion of ELP flares varies greatly because only very few flares occurred each month. However, once a considerable number of flares occurred (starting in February 2011), the percentage of ELP flares among them stabilized.\\
\indent Next, we compare the number and percentage of ELP flares with the 13-month smoothed monthly mean sunspot number.\footnote{Source: WDC--SILSO, Royal Observatory of Belgium, Brussels; data available at \url{https://www.sidc.be/silso/datafiles}.} To this extent, we also applied a running mean to the ELP data analogously to how the 13-month smoothed sunspot data are calculated. Figure~\ref{fig:elp_vs_cycle} shows the smoothed sunspot numbers along with the smoothed number of ELP flares (top panel) and the percentage of ELP flares calculated from the smoothed values (bottom panel). The number of ELP flares rises from solar minimum to the first maximum, stays approximately constant for a couple of months, and subsequently decreases again during the Gnevyshev gap in-between the sunspot cycle maxima. The percentage of ELP flares stays roughly constant during the first rise time of the cycle, followed by a substantial decrease during the first maximum. Overall, we find a modest to strong positive correlation (\mbox{$P_c = 0.60$}, \mbox{$CI_{95\%} = [0.38,0.76]$}) between the 13-month smoothed number of ELP flares and the 13-month smoothed monthly mean total sunspot number. When considering the percentage of ELP flares, we find a strong negative correlation (\mbox{$P_c = -0.75$}, \mbox{$CI_{95\%} = [-0.86,-0.59]$}) with the sunspot number.


\subsubsection{ELP flares and eruptivity}\label{subsec:erup_results}

\begin{figure}[h!]
    \centering
    \includegraphics[width=0.45\textwidth]{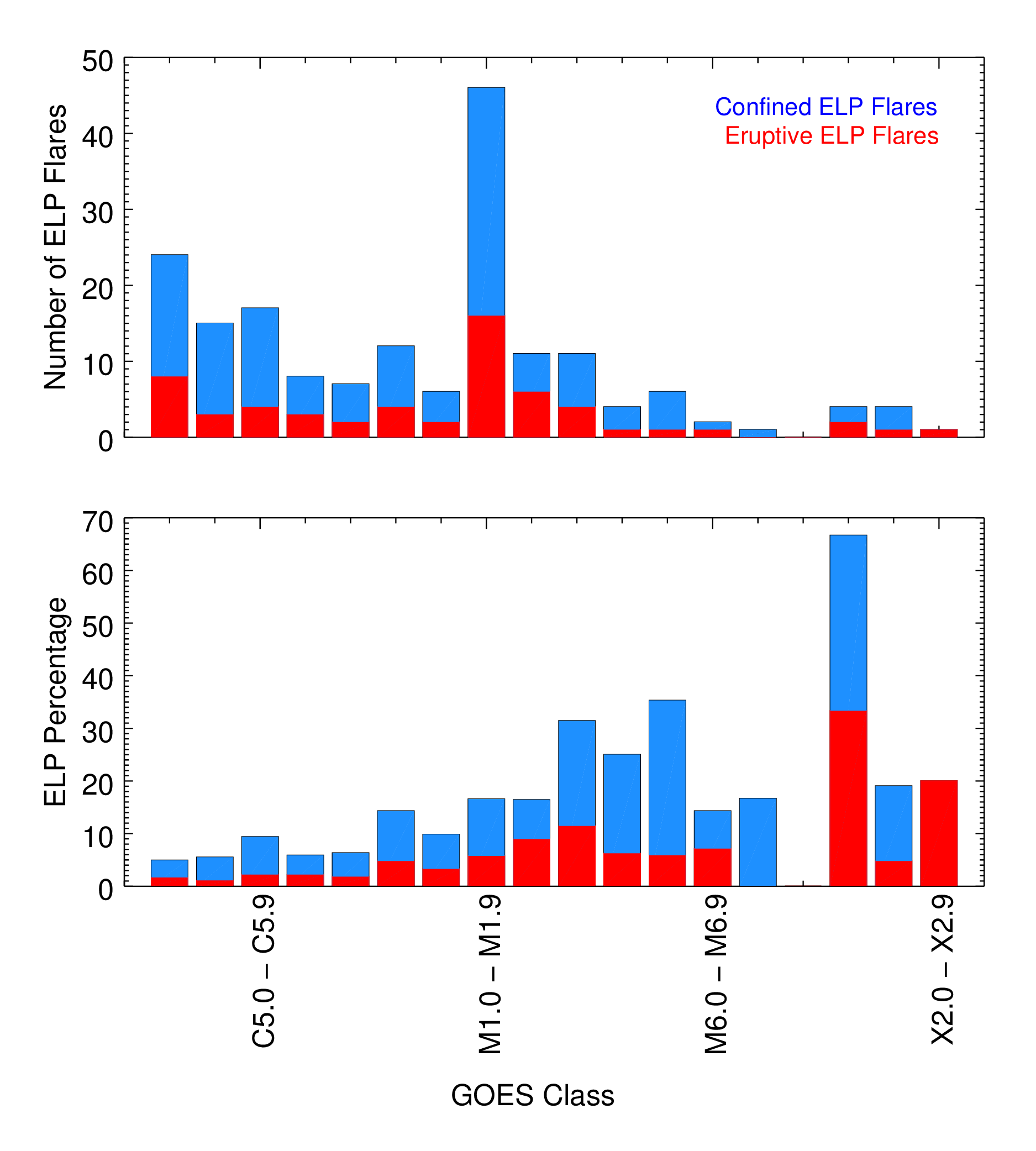}
    \caption{Distribution of ELP flares as a function of GOES SXR flare class. Top panel: Absolute number of ELP flares. Bottom panel: ELP flare percentage. Confined ELP flares are shown in blue, eruptive flares in red. The combined length of the red and blue bars for each bin signifies the total number (top panel) or percentage (bottom panel) in this bin.}
    \label{fig:flares_per_goes_class}
\end{figure}
\begin{figure}
    \centering
    \includegraphics[width=0.4\textwidth]{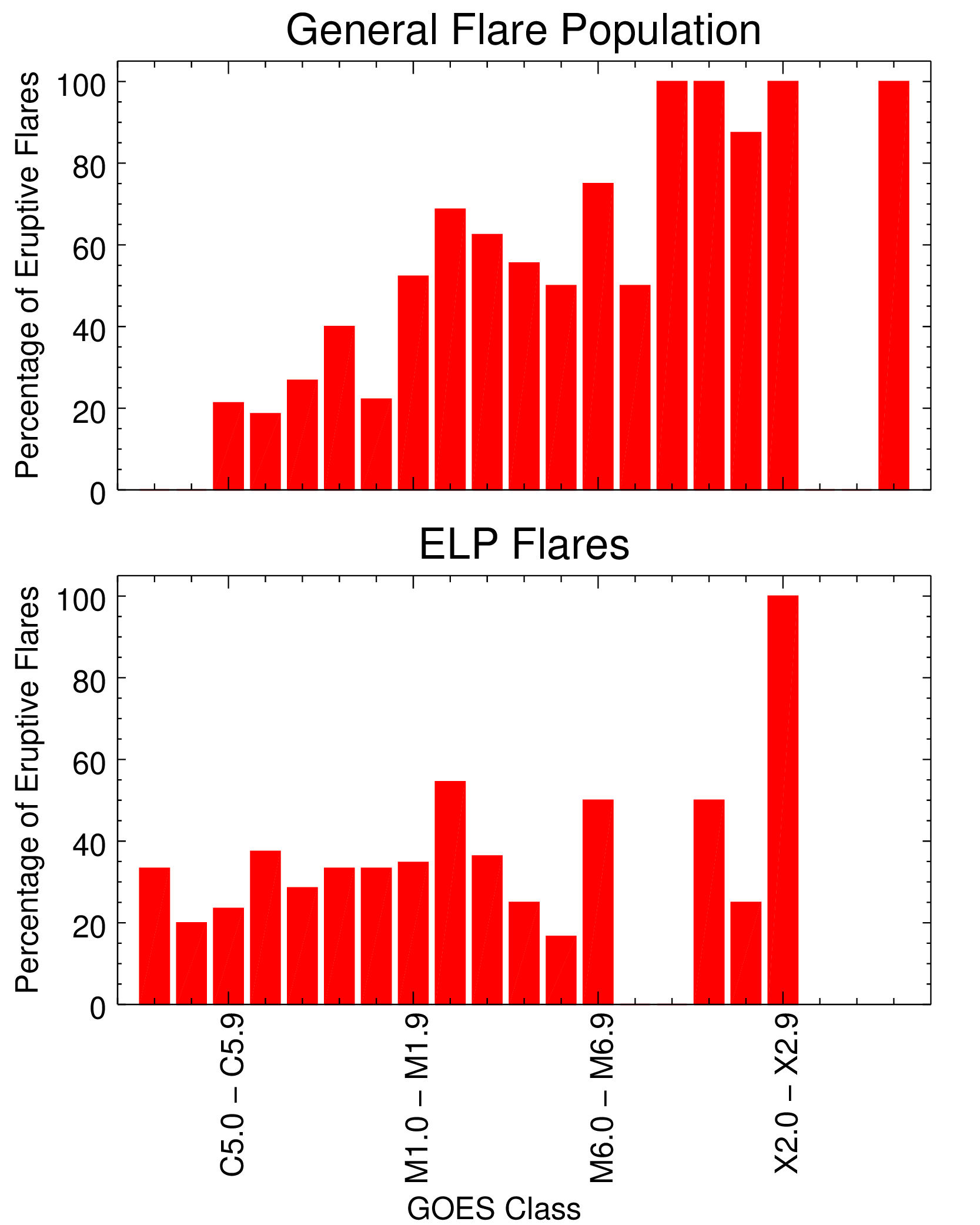}
    \caption{Percentage of eruptive flares as a function of GOES SXR class. Top panel: Flares included in the subset of \cite{2021ApJ...917L..29L} (i.e., flares $\geq$C5.0 between June 2010 and September 2017). Bottom panel: ELP flares.}
    \label{fig:erup_flares_per_goes_class}
\end{figure}
C-class flares make up the largest absolute number of ELP flares, while larger flares (M- and X-class) have a higher percentage of flares identified as ELPs (see Table~\ref{tab:elp_classes}). A more detailed breakdown of GOES-class distribution is given in Fig.~\ref{fig:flares_per_goes_class}, which shows the absolute number of ELP flares as a function of the GOES class in its top panel and the corresponding percentage of all flares identified as ELPs in its bottom panel. Both panels show the values separately for eruptive and confined ELP flares. A total of 46 ELP flares have a GOES class between M1.0 and M1.9, almost double the next-highest peak of 24 flares between C3.0 and C3.9. In general, the percentage of ELP flares seems to increase for larger flare classes. The median flare class is M1.0 for ELPs and C5.7 for all flares, respectively.\\
\indent Overall, of the 179 ELP flares observed during the study period, 59 (33\%) are identified as eruptive and 120 flares (67\%) as confined. Due to the occasional uncertainty in establishing a connection between flares and CMEs (see Sect.~\ref{subsubsec:erup}), not every event is clearly eruptive or certainly confined. For 8 events it is unclear whether they have an associated CME (they are indicated by a star in the flare list, see Sect.~\ref{sec:data_av} and Appendix~\ref{sec:flare_list}).\\
\indent In Fig.~\ref{fig:erup_flares_per_goes_class} we compare the GOES class distribution of the eruptivity of ELP flares with those of the general flare population. The top panel displays the percentage of eruptive flares among all flares $\geq$C5.0 in the subset studied in \cite{2021ApJ...917L..29L} for that time period, while the bottom panel illustrates the same for ELP flares. A summary of all eruptive and confined (ELP) flares is given in Table~\ref{tab:eruptivity_of_elps_class}. The percentage of eruptive flares is strikingly lower for ELP flares than for the general flare population. Every single bin from M1.0--M1.9 to X1.0--X1.9 shows a lower eruptivity by double-digit percentage points. On the other hand, ELP flares between C5.0 and C9.9 are more likely to be eruptive than their classical counterparts. In total, C-class ELP flares have a higher likelihood of being associated with an eruption (by 4.4 percentage points) than the general flare population, while M- and X-class ELPs have a significantly lower CME-association rate (by 21.7 and 50.9 percentage points, respectively).\\
\begin{table}
    \caption{Number and percentage of eruptive (subscript $er$) and confined (subscript $cf$) flares $\geq$C5.0 (top half), and ELP flares (bottom half) between May 1, 2010, and May 26, 2014, for GOES SXR class C, M, and X, respectively.}\label{tab:eruptivity_of_elps_class}
    \centering
    \begin{tabular}{ |c|c|c|c|c|c| }
    \hline
    \multicolumn{6}{|c|}{\textbf{All Flares $\geq$C5.0}} \\
    \hline
    \specialcell{GOES\\Class} & $\mathrm{n}$ & $\mathrm{n_{er}}$ & $\mathrm{n_{cf}}$ & \specialcell{$\mathrm{\frac{n_{er}}{n}}$ (\%)} & \specialcell{$\mathrm{\frac{n_{cf}}{n}}$ (\%)}\\
    \hline
    C & 242 & 60 & 182 & 24.8 & 75.2 \\
    M & 184 & 107 & 77 & 58.2 & 41.8 \\
    X & 11 & 10 & 1 & 90.9 & 9.1 \\
    \hline
    Sum & 437 & 177 & 260 & 40.5 & 59.5 \\
    \hline
    \multicolumn{6}{|c|}{\textbf{ELP Flares}} \\
    \hline
    \specialcell{GOES\\Class} & $\mathrm{n_{ELP}}$ & $\mathrm{n_{ELP,er}}$ & $\mathrm{n_{ELP,cf}}$ & \specialcell{$\mathrm{\frac{n_{ELP,er}}{n_{ELP}}}$ (\%)} & \specialcell{$\mathrm{\frac{n_{ELP,cf}}{n_{ELP}}}$ (\%)}\\
    \hline
    C & 89 & 26 & 63 & 29.2 & 70.8 \\
    M & 85 & 31 & 54 & 36.5 & 63.5 \\
    X & 5 & 2 & 3 & 40.0 & 60.0 \\
    \hline
    Sum & 179 & 59 & 120 & 33.0 & 67.0 \\
    \hline
    \end{tabular}
    \tablefoot{The first column labels the GOES class that the flares belong to. The second column signifies the total number of flares in the corresponding class. The third and fourth columns depict the number of eruptive and confined flares, while column 5 and 6 show the corresponding percentage of eruptive and confined flares with regards to the overall number of flares in that class. The eruptivity data for the flares $\geq$C5.0 are from \cite{2021ApJ...917L..29L}.}
\end{table}
\begin{figure}
    \centering
    \includegraphics[width=0.47\textwidth]{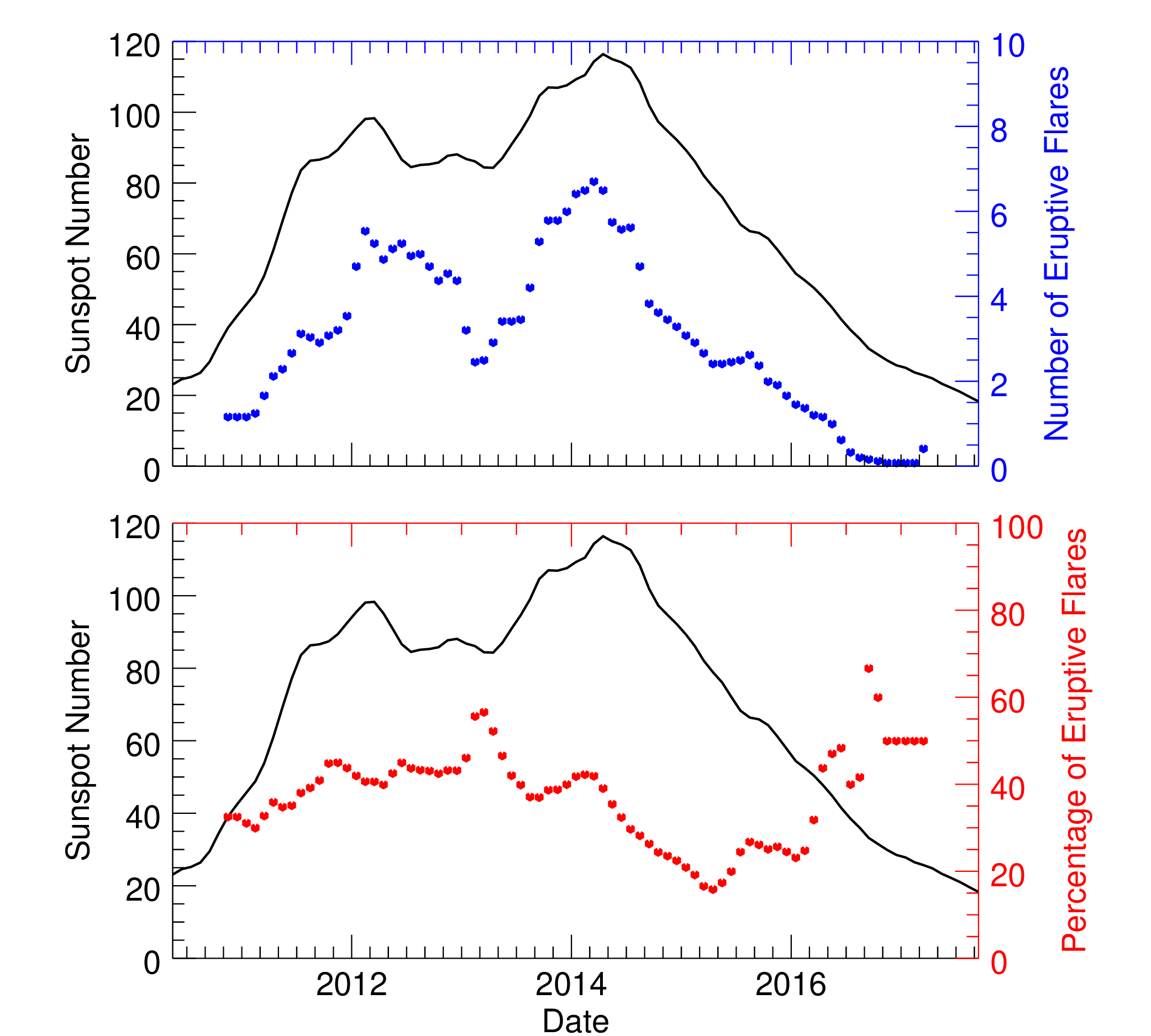}
    \caption{13-month smoothed monthly total number (top) and percentage (bottom) of eruptive flares relative to the overall number of flares $\geq$C5.0 compared to the 13-month smoothed monthly mean total sunspot number (black curve) as a function of time between June 2010 and September 2017. The eruptivity data for the flares are from \cite{2021ApJ...917L..29L}.}
    \label{fig:erup_vs_cycle}
\end{figure}
\indent Since ELP flares tend more toward being confined than the general flare population, and since Fig.~\ref{fig:elp_vs_cycle} has shown a drop in the relative share of ELPs toward solar maximum, we used the data from \cite{2021ApJ...917L..29L} to compute the eruptivity as a function of time from June 2010 to September 2017 to investigate whether the change in eruptivity of flares could be responsible for the lower probability of ELP flares at high solar activity. We again applied a 13-month running mean. We find an unusually high eruptivity of solar flares toward the minimum at the end of the solar cycle (Fig.~\ref{fig:erup_vs_cycle}, top panel). Furthermore, we see that during the rise and fall times of the solar cycle, the percentage of eruptive flares evolves very similarly (albeit in opposite directions). Overall, we find a strong correlation between the 13-month smoothed number of eruptive flares (in the general population) and the 13-month smoothed monthly mean sunspot number ($P_c = 0.94$, $CI_{95\%} = [0.91,0.95]$). However, looking at the percentage of eruptive flares, the correlation is almost nonexistent (Fig.~\ref{fig:erup_vs_cycle}, bottom panel), at $P_c = -0.18$ ($CI_{95\%} = [-0.38,-0.04]$).


\subsection{ELP parameters and relation to the main flare} \label{subsec:elp_params_results}


\subsubsection{Relative strength of the EUV late phase}\label{subsec:peak_ratio_results}

\begin{figure}
    \centering
    \includegraphics[width=0.42\textwidth]{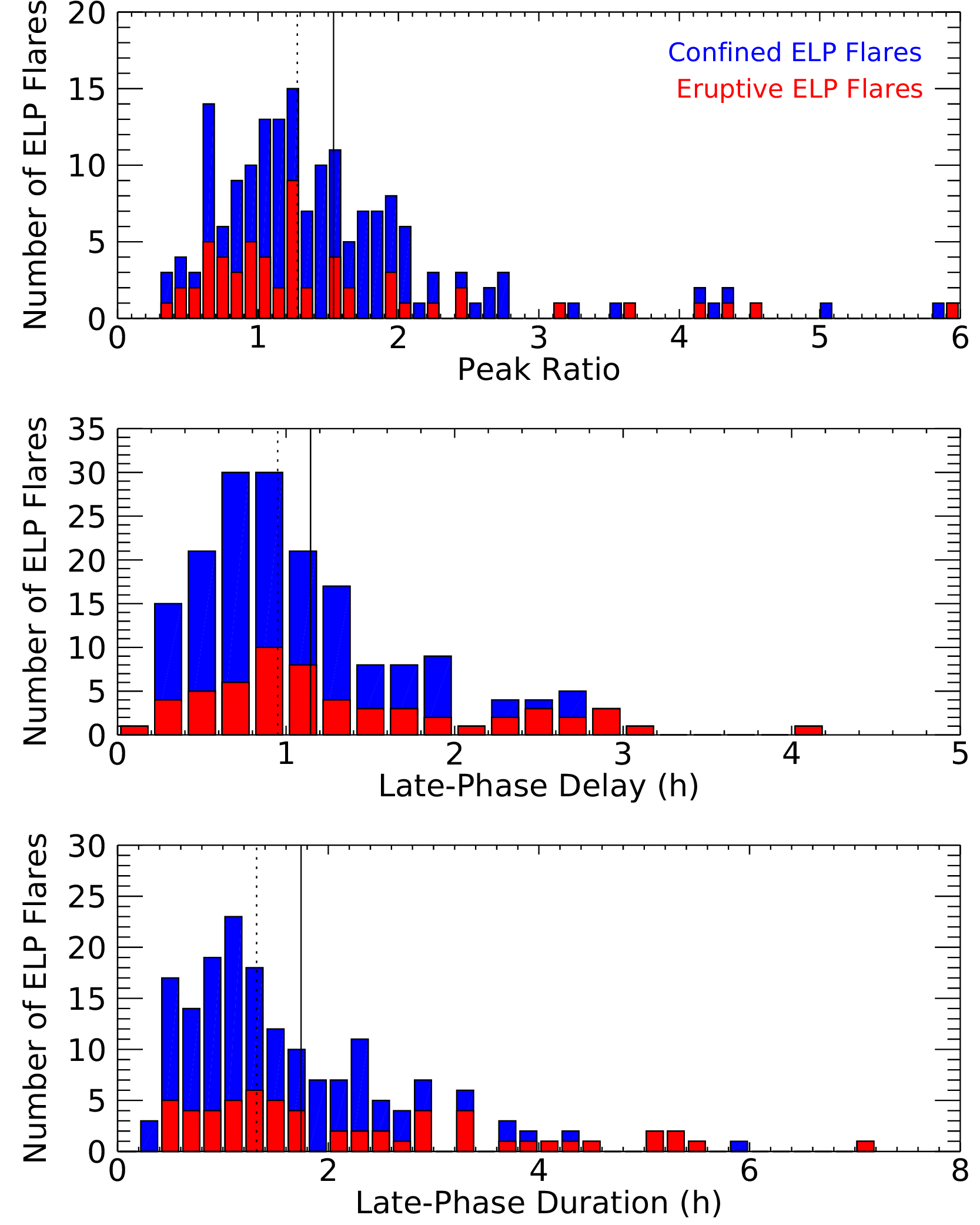}
    \caption{Histograms of the peak ratio (top), the late-phase delay (middle), and the late-phase duration (bottom) derived from the EVE~33.5~nm measurements for all ELP flares. Eruptive ELP flares are shown in red, confined flares in blue. The mean and median of the distributions are shown as solid and dotted vertical black lines, respectively.}
    \label{fig:lp_params_hist}
\end{figure}

For the ELP flares presented in this work, the peak ratio of the late-phase to the main-phase peak in the EVE Fe~XVI~33.5~nm irradiance ranges between 0.3 and 5.9, with a mean (median) of $1.54\pm0.97$ ($1.28\pm0.43$). Figure~\ref{fig:lp_params_hist} (top panel) shows the distribution of the peak ratios. Of the 179 ELP events, 128 (71.5\%) can be classified as extreme ELP events \citep{2015ApJ...802...35L} defined by a peak ratio $>$1, while only 49 (27.4\%) have a peak ratio below 1. Two flares have a main maximum with a PFS irradiance value below 0 (probably due to occultations close to the limb), therefore inhibiting the calculation of a meaningful peak ratio.
\begin{figure}
    \centering
    \includegraphics[width=0.47\textwidth]{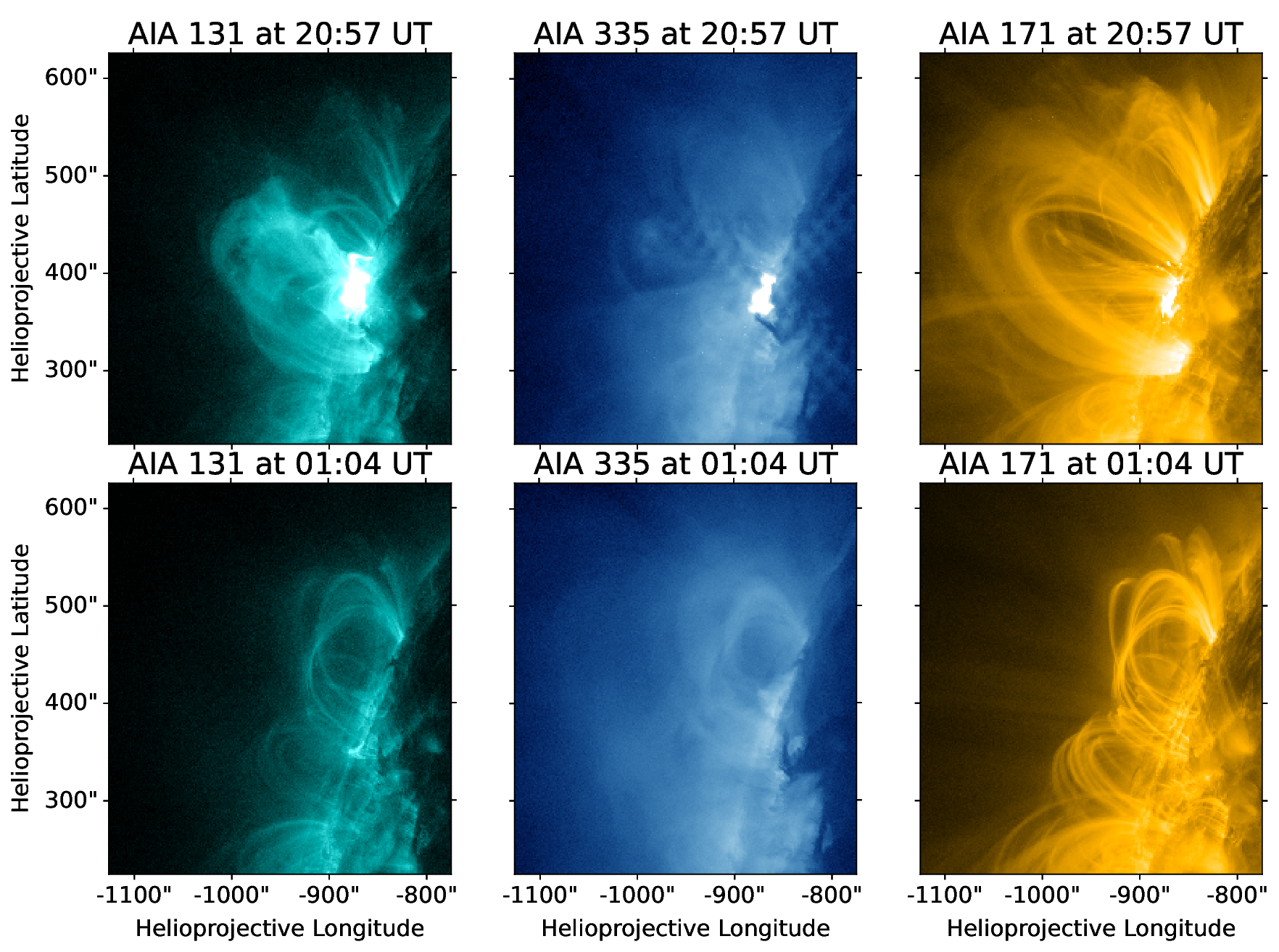}\\
    \includegraphics[width=0.47\textwidth]{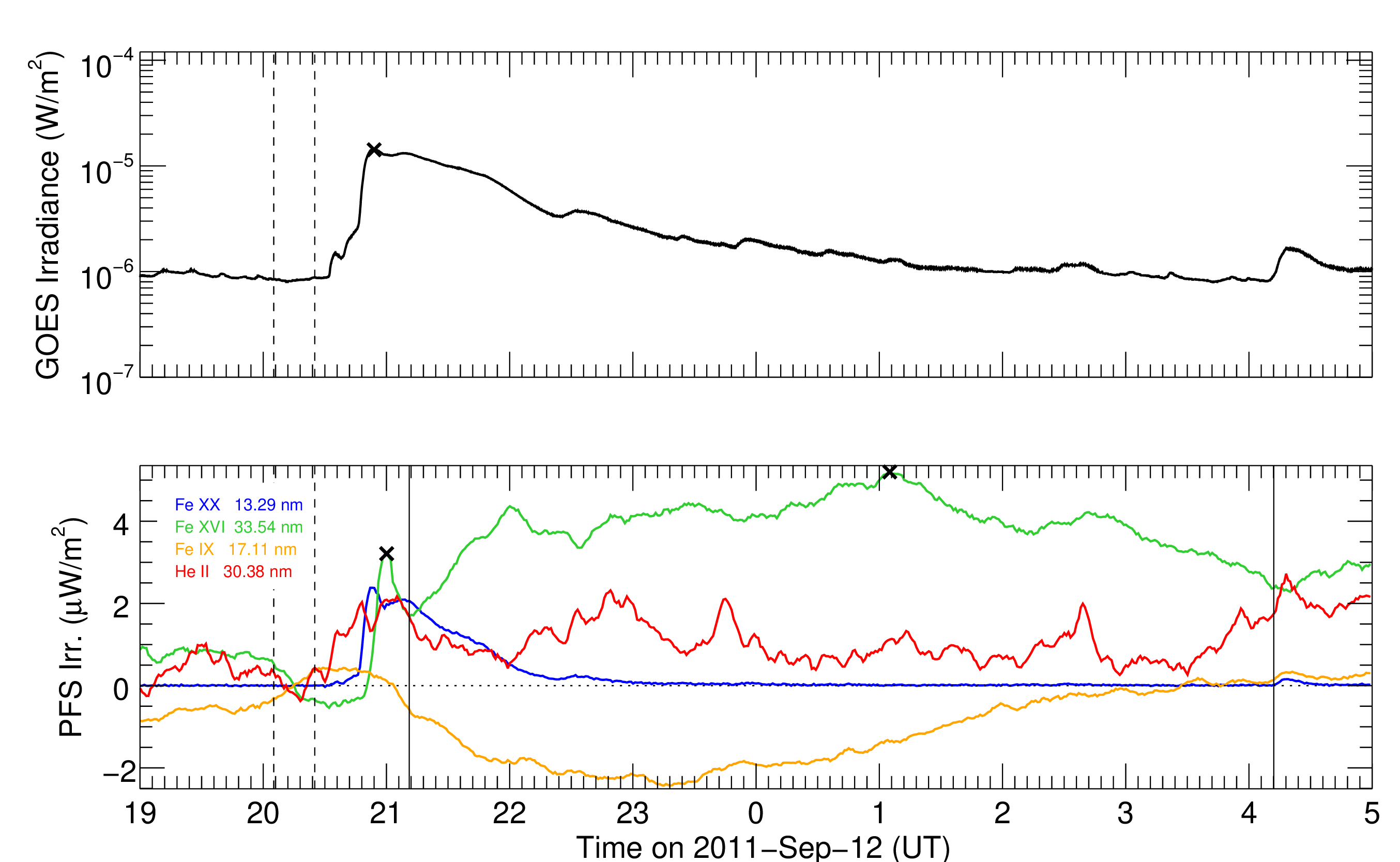}
    \caption{Same as Fig.~\ref{fig:aia_eve_high_ratio}, but for the eruptive C9.9~flare on September~12,~2011.}
    \label{fig:aia_eve_long_dur}
\end{figure}


\subsubsection{ELP delay and duration}\label{subsubsec:del_dur_relation_results}

The bottom two panels of Fig.~\ref{fig:lp_params_hist} show the distributions of the late-phase delay and its duration for all ELP flares. The late-phase delay ranges between 12~minutes and 245~minutes, with a mean of $82\pm41$~minutes and a median of $57\pm21$~minutes. The late-phase duration spans from 22~minutes to 421~minutes, with a mean and median of $104\pm73$~minutes and $79\pm37$~minutes, respectively. The results are summarized in Table~\ref{tab:lp_params_results}.\\
\indent Figure~\ref{fig:aia_eve_long_dur} shows the GOES and EVE light curves as well as AIA images for the ELP flare with the longest duration (the eruptive C9.9 flare on September 12, 2011). The times were chosen in order to highlight the main-flare and the late-phase emissions. The three-part structure of the associated CME is visible in the AIA 33.5~nm images in the top row. The figure and the associated movie show that the combination of cooling plasma (seen as bright regions in the center left panel during the main-flare peak and in the bottom center panel during the late-phase peak) and energy release during the late phase (newly formed loops can be observed in the bottom row of panels) seems to be responsible for the extended EUV late phase.\\
\begin{table}
    \centering
    \caption{Statistical properties of the late-phase delay ($\mathrm{\Delta t}$), the late-phase duration (T$_{\mathrm{LP}}$) and the peak ratio for all ELP flares (top), as well as for eruptive (middle) and confined ELPs (bottom).}\label{tab:lp_params_results}
    \begin{tabular}{ |c|c|c|c|c|c| }
    \hline
    \multicolumn{6}{|c|}{\textbf{All ELP Flares}} \\
    \hline
    Property & Mean & $\mathrm{\sigma}$ & Median & Min. & Max.\\
    \hline
    $\mathrm{\Delta t}$ & 1.15 & 0.69 & 0.95 & 0.20 & 4.08 \\
    T$_{\mathrm{LP}}$ & 1.74 & 1.22 & 1.32 & 0.37 & 7.02 \\
    Peak Ratio & 1.54 & 0.97 & 1.28 & 0.30 & 5.90 \\
    \hline
    \multicolumn{6}{|c|}{\textbf{Eruptive ELP Flares}} \\
    \hline
    Property & Mean & $\mathrm{\sigma}$ & Median & Min. & Max.\\
    \hline
    $\mathrm{\Delta t}$ & 1.37 & 0.87 & 1.17 & 0.20 & 4.08 \\
    T$_{\mathrm{LP}}$ & 2.21 & 1.56 & 1.67 & 0.42 & 7.02 \\
    Peak Ratio & 1.48 & 1.11 & 1.24 & 0.31 & 5.90 \\
    \hline
    \multicolumn{6}{|c|}{\textbf{Confined ELP Flares}} \\
    \hline
    Property & Mean & $\mathrm{\sigma}$ & Median & Min. & Max.\\
    \hline
    $\mathrm{\Delta t}$ & 1.03 & 0.56 & 0.88 & 0.30 & 2.77 \\
    T$_{\mathrm{LP}}$ & 1.51 & 0.93 & 1.25 & 0.37 & 5.92 \\
    Peak Ratio & 1.56 & 0.90 & 1.40 & 0.30 & 5.80 \\
    \hline
    \end{tabular}
    \tablefoot{The values for the delay and duration are given in hours.}
\end{table}
\begin{figure}
    \centering
    \includegraphics[width=0.49\textwidth]{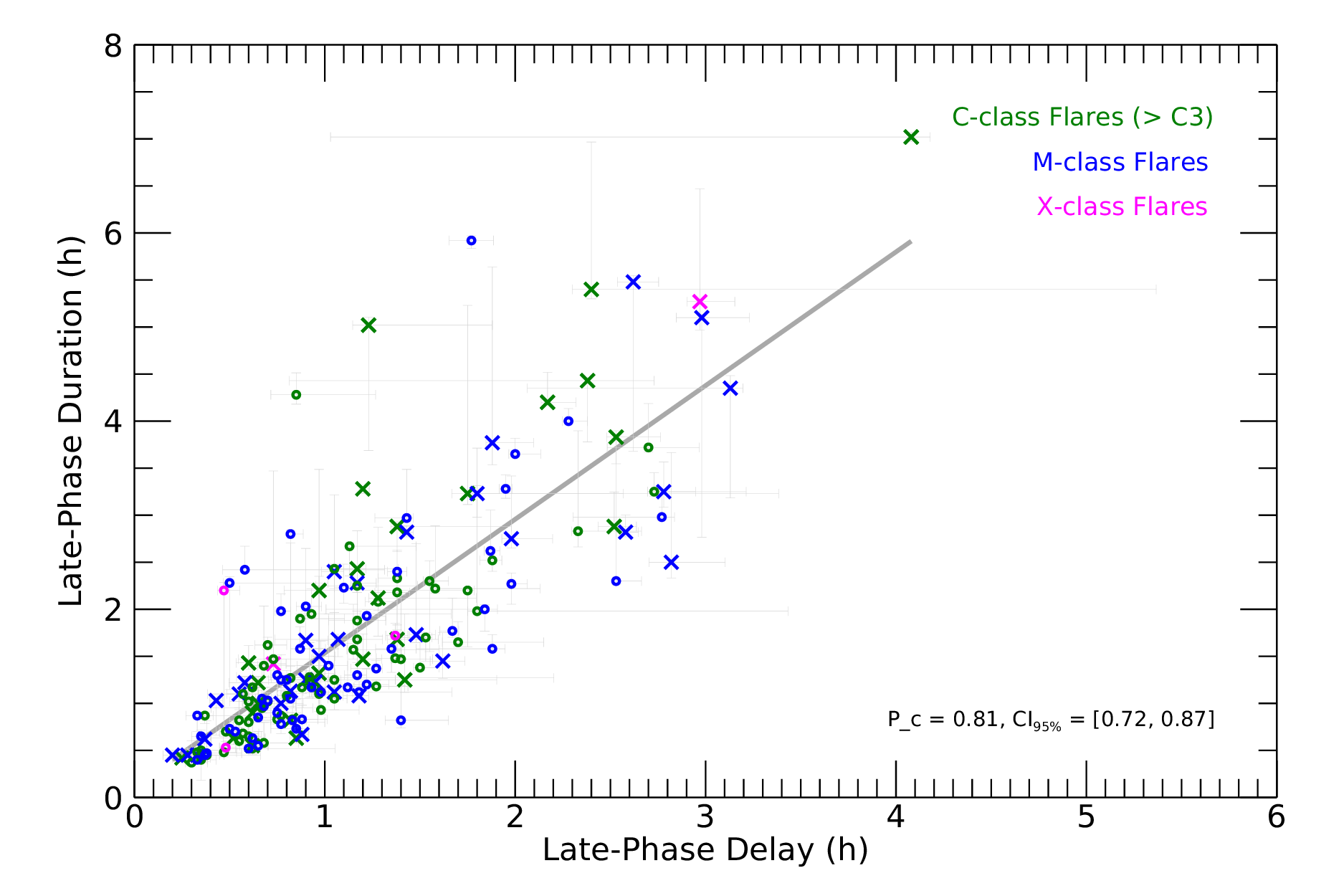}
    \caption{Duration of the late phase as a function of the time delay from the main phase. C-class flares are shown in green, M-class flares in blue, and X-class flares in magenta. Eruptive flares are marked with a cross, confined flares with a circle. The error bars are shown as thin gray lines. A linear fit is given as a solid gray line.}
    \label{fig:lp_del_lp_dur}
\end{figure}
\indent Figure~\ref{fig:lp_del_lp_dur} shows the EUV late-phase duration of all ELP flares as a function of the EUV late-phase delay. The scatter plot reveals a strong positive correlation between the two parameters (\mbox{$P_c = 0.81$}, $CI_{95\%} = [0.72,0.87]$). This correlation is slightly stronger for eruptive flares ($P_c = 0.87$, \mbox{$CI_{95\%} = [0.74,0.93]$}) than for confined ones (\mbox{$P_c = 0.71$}, \mbox{$CI_{95\%} = [0.57,0.81]$}). Applying a linear fit to the data, we find the fit coefficients to be $a = 0.115$ ($CI_{a,95\%} = [-0.072,0.307]$) for the $y$-intercept and $b = 1.421$ ($CI_{b,95\%} = [1.233,1.605]$) for the slope. The linear fit is depicted as a thick gray line in Fig.~\ref{fig:lp_del_lp_dur}.


\subsection{ELP flare-ribbon morphology} \label{subsec:ribbons_results}

66 of the 179 ELP flares were either located too close to the solar limb, making a reliable determination of the flare-ribbon morphology impossible due to the strong projection effects, or appeared at a time when AIA images were not available (e.g., early May 2010). This still leaves almost two thirds of the flares (113) where the ribbons were well visible and close enough to the solar center so that projection effects did not distort them too much. The largest share of ELP flares produced complex (irregular) ribbons, with 42 flares belonging to this group (37.2\%). Similarly, the classical, more or less parallel, pair of elongated ribbons make up a comparable portion, at 39 flares (34.5\%). Circular ribbons are the least common for ELP flares, since only 32 flares (28.3\%) created this type of ribbons. In Table~\ref{tab:ribb_year_class} we summarize the results separated by flare class and eruptivity. It seems that C-class ELP flares are slightly more likely to produce the classical system of two ribbons, whereas this preference slightly shifts toward a more complex configuration for M-class ELPs. In terms of eruptivity, circular-ribbon ELP flares are most likely to be associated with a CME (31.3\%), whereas complex-ribbon ELP flares have a substantially lower likelihood of being eruptive (21.4\%). We do not find a distinct relation to the ELP strength: 30 two-ribbon (77\%), 28 complex-ribbon (67\%), and 20 circular-ribbon ELP flares (63\%) had their late-phase peak exceed the main-phase peak.\\
\begin{table}
    \caption{Number of ELP flares across the three types of flare-ribbon morphology as a function of GOES class and eruptivity.}\label{tab:ribb_year_class}
    \centering
    \begin{tabular}{ |c|c|c|c| }
    \hline
    Flare Class & Two & Circular & Complex \\
    \hline
    C & 20 & 13 & 19 \\
    M & 19 & 17 & 22 \\
    X & 0 & 2 & 1 \\
    \hline
    Eruptivity & Two & Circular & Complex \\
    \hline
    Eruptive & 10 & 10 & 9 \\
    Confined & 29 & 22 & 33 \\
    \hline
    Sum & 39 & 32 & 42 \\
    \hline
    \end{tabular}
    \tablefoot{The first column gives the respective class, while columns 2--4 indicate the number of two-, circular-, and complex-ribbon ELP flares, respectively.}
\end{table}


\subsection{Properties of ELP-productive active regions} \label{subsec:ar_results}

\begin{figure}
    \centering
    \includegraphics[width=0.41\textwidth]{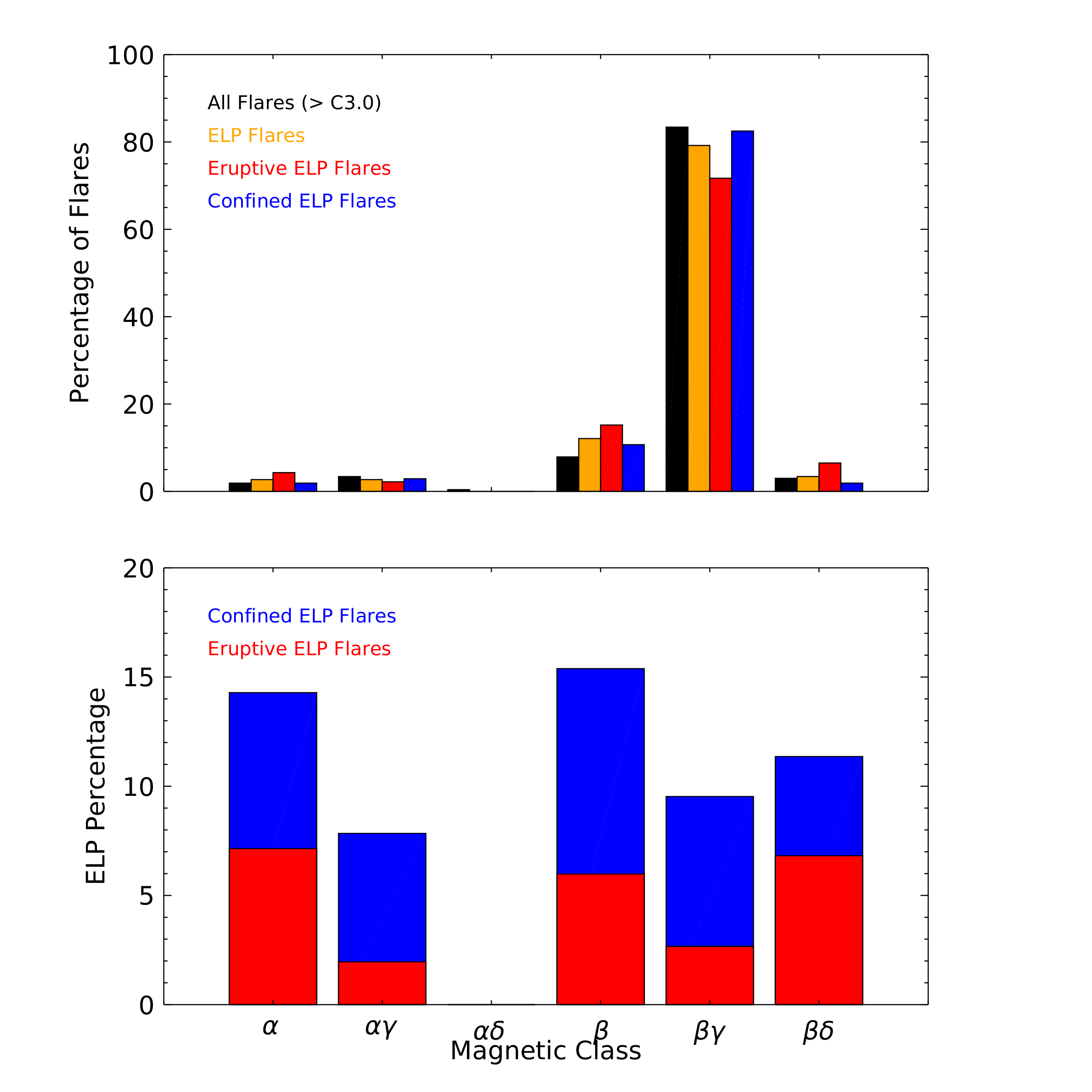}
    \caption{Relative number of flares (top) as well as percentage of flares identified as ELPs (bottom) as a function of the Mount Wilson magnetic classification of the flare-harboring sunspot region for all flares with a GOES class exceeding C3.0 (black), all ELP flares (orange), as well as the eruptive (red) and confined (blue) portion of ELP flares. The y-axis in the top panel displays the share of flares belonging to the respective group divided by the total number of flares in this colored category (in~\%).}
    \label{fig:mag_class}
\end{figure}

99 ARs produced ELP flares, while 211 flare-productive ($\geq$C3.0) ARs did not exhibit any ELP-flare activity. That means a significant portion --- about a third --- of ARs produced at least one ELP flare during their lifetime. 8 ELP flares are not associated with an AR (7 of them due to their appearance over the limb, while the remaining one occurred between two ARs). The vast majority of ELP-productive ARs exhibited three or less ELPs. AR 11339 was the most productive, with 8 ELP flares (out of a total of 30 flares $\geq$C3.0 created during its lifetime). Another notable AR is AR 11598, which produced 9 flares ($\geq$C3.0) overall, 6 of which were ELPs (4 confined, 2 eruptive). Table~\ref{tab:elp_prod} shows how ELP productivity is partitioned across ELP-productive ARs.\\
\begin{table}
    \caption{ELP-flare productivity of active regions.}\label{tab:elp_prod}
    \centering
    \begin{tabular}{ |c|c|c| }
    \hline
    No. ELPs & No. ARs & Relative Share (\%)\\
    \hline
    1 & 63 & 58.9 \\
    2 & 18 & 16.8 \\
    3 & 9 & 8.4 \\
    4 & 4 & 3.7 \\
    5 & 3 & 2.8 \\
    6 & 1 & 0.9 \\
    7 & 0 & 0 \\
    8 & 1 & 0.9 \\
    n.a. & 8 & 7.5 \\
    \hline
    \end{tabular}
    \tablefoot{The first column describes the number of ELPs that were generated by the ARs. Here, ``n.a.'' means that the flares could not be associated with an AR. The second column denotes the number of ARs that produced the corresponding number of ELPs. The last column describes the relative share of ARs of this productivity relative to the total number of ELP-productive ARs. Only ARs that produced at least one ELP flare are included in this table.}
\end{table}
\indent Figure~\ref{fig:mag_class} shows the distribution of ELP flares across the Mount Wilson AR magnetic classes. One can see that the complex $\beta\gamma$ class is represented less in ELP flares than in the general flare population, while the simpler $\beta$ configuration is more common. Analogously, we see a shift toward the simple $\alpha$ structure, rather than the more complex structures (e.g., $\alpha\gamma$). This shift toward simpler configurations is not uniform across all ELP flares: Confined ELP flares align more closely with the general flare population. The distribution of ELP-flaring ARs across McIntosh classes shows a similar increase in simpler configurations (see Fig.~\ref{fig:mcintosh_results}). However, only class Ekc deviates significantly from the general flare population, being considerably less likely to underlie ELP flares.


\section{Discussion}\label{sec:discussion}


This study has shown that roughly 10\% of all flares $\geq$C3.0 between May 2010 and May 2014 can be classified as ELP flares. This is in agreement with the study by \cite{2011ApJ...739...59W}, who classified 13\% of their 191 flares as ELPs. The percentage of flares identified as ELPs shows a negative correlation with the sunspot number, which is consistent with the findings of \cite{2014SoPh..289.3391W}. Two thirds of ELP flares are confined, while one third is associated with an eruption; a fraction similar to the general flare population and comparable to the results put forth by \cite{2020ApJ...890..158C}, who found 40\% of the 55 flares included in their study to have been eruptive. The late-phase peak shows a higher intensity enhancement than the main-phase peak in approximately 70\% of ELP flares, a considerably higher fraction than previously reported (cf. \cite{2020ApJ...890..158C}). We find a strong linear relationship between the delay of the late phase and its duration (\mbox{$P_c = 0.81$}, $CI_{95\%} = [0.72,0.87]$). In the following we discuss some of these aspects in more detail.\\
\indent Both the late-phase delay and duration are notably longer for eruptive than for confined flares (by 33\% and 15\%, respectively). This is possibly a result of the creation mechanism, assuming that the late-phase loops in confined cases are heated at (or at least close to) the same time as the main-phase loops, and assuming that eruptive late-phase loops get created via reconnection some time after the eruption (and therefore after the main flare). The maximum value of the late-phase delay for confined flares found in Sect.~\ref{subsubsec:del_dur_relation_results} is 180~minutes. Since the cooling time is a function of loop length, this puts an upper limit on the length of late-phase loops in confined cases (if we assume they are heated at the same time as the main-phase loops and have the same density). Using the formula for the cooling time of a flare loop derived by \mbox{\cite{1995ApJ...439.1034C}},
\begin{equation}\label{eq:8_cooling_time}
    \tau_{cool} \doteq 2.35 \times 10^{-2} \, L^{5/6} \, T_e^{-1/6} \, n_e^{-1/6},
\end{equation}
we can get a basic estimate for the loop half-length,
\begin{equation}\label{eq:9_loop_half_length}
    L \approx 90 \, \tau_{cool}^{6/5} \, T_e^{1/5} \, n_e^{1/5},
\end{equation}
where $\tau_{cool}$ is the cooling time and $T_e$ and $n_e$ are the electron temperature and electron density at the beginning of the cooling, respectively. A detailed analysis of the flares with long cooling times would need to be carried out to determine the necessary parameters, which is outside the scope of this work. Here, we try a basic estimate of the mean loop half-length of confined ELP flares by assuming \mbox{$T_e = 1.0 \times 10^7$ K} (i.e., the temperature observed in AIA 131~\AA, in order to be able to compare with observational estimates) and \mbox{$n_e = 10^{10}$ cm$^{-3}$}, which are realistic values for a late-phase flare arcade following \cite{2002ESASP.505..207P} and \mbox{\cite{2013ApJ...768..150L}}. Assuming uncertainties in the temperature, density, and cooling time of $3$~MK, $5 \times 10^{9}$~cm$^{-3}$, and $5$~minutes, respectively, we arrive at an estimated maximum loop half-length for confined ELP flares of $L = 254 \pm 55$~Mm. If we insert the mean value of the late-phase delay for confined flares (62 minutes), we obtain $L = 71 \pm 19$~Mm. This is similar to the loop half-lengths found for the two ELP flares discussed in \mbox{\cite{2013ApJ...768..150L}}. Doing the same exercise for the flares shown in Fig.~\ref{fig:aia_eve_high_ratio} and Fig.~\ref{fig:aia_eve_long_dur}, we obtain $L = 102 \pm 26$ and $L = 368 \pm 70$~Mm, respectively. We note that these values are significantly higher than a simple estimate of the loop half-length from AIA images gives ($L = 76 \pm 11$ and $L = 260 \pm 11$~Mm), but the relative changes between the two are roughly fitting.\\
\indent The strong positive correlation between the late-phase delay and the late-phase duration (Fig.~\ref{fig:lp_del_lp_dur}, \mbox{$P_c = 0.81$}) can be explained via the cooling timescale introduced above, which is a function of the loop length. Therefore, longer loops take longer to cool down, which means that the delay until they reach warm coronal temperatures (observed in the 33.5~nm wavelength range) is longer for larger loops. In turn, the loops also spend more time at this temperature, creating an approximately linear relationship between the late-phase delay and its duration (which will be further promoted by slightly different initial conditions for the individual strands of the coronal loops). This relation is even stronger for eruptive (\mbox{$P_c = 0.87$}) than confined flares (\mbox{$P_c = 0.71$}), which can be the result of additional heating during the late phase in some confined cases, increasing the duration of the late phase relative to its delay from the main peak.\\
\indent The majority of ELP flares found in this work (128 out of 179, or 71.5\%) can be classified as extreme ELP events (i.e., the peak ratio is $>$1). 61\% of eruptive ELP flares had their late-phase peak exceed the first peak, while 77\% of confined flares did so, which shows the same tendency as previously reported \citep{2016ApJS..223....4W,2020ApJ...890..158C}. Similarly to the findings reported by \cite{2020ApJ...890..158C}, two-ribbon flares have a higher probability of producing extreme ELP events (77\%) than flares with circular (63\%) or complex ribbons (67\%). However, all of these results show a much higher share than previously reported by \cite{2020ApJ...890..158C}, where only 40\% of all ELP flares identified by the authors had such high late-phase peaks. The reasons for this are the following: (1) The biggest reason is that we used the smoothed PFS irradiance values (as described in Sect.~\ref{subsec:data_prods_and_prep}), while \cite{2020ApJ...890..158C} did not apply any smoothing; in general, smoothing considerably lowers the (usually spikier) main peak compared to the (usually broader) late-phase peak, as can be seen in Fig.~\ref{fig:eve_raw}, which makes comparisons with other studies that did not use such smoothing difficult, since it generally alters the main--to--late-phase peak ratio, (2) some of the (low-ratio) ELP flares in that paper are not classified as such according to our criteria, which is mainly due to our criteria 3 and 4 (see Sect.~\ref{subsec:elp_crit}), and (3) the flare pre-selection process described in Sect.~\ref{subsec:elp_crit} might result in some low-ratio ELP flares being missed, since the time series plots available in the EVE flare catalog of \cite{2012PhDT........57H} are scaled to the highest peaks within a nine-hour time window, making it hard to determine possible late-phase candidates when there is at least one major flare (e.g., X-class) happening within this time frame.\\
\indent The absolute number of ELP flares is not uniform across GOES classes. M-class flares are most likely to produce an ELP event (at 19.3\%), followed by X-class flares (16.1\%) and C-class flares (6.7\%). These values follow the trend found by \cite{2014SoPh..289.3391W}, who also noted the higher likelihood of ELP flares for M- and X-class flares compared to the lower-energy C-class flares. Nonetheless, the probability for class M and X determined in this work is still way below the 50\% reported by \cite{2014SoPh..289.3391W}, who used the dual-decay behavior of the GOES SXR of flares as a proxy for ELPs.\\
\indent The Mount Wilson magnetic class of sunspot regions shows that ELP flares --- and especially eruptive ones --- exhibit a tendency toward simpler magnetic configurations compared to the general flare population (Fig.~\ref{fig:mag_class}). This is in stark contrast to previous findings \citep{2011ApJ...739...59W}, and also contradicts the fact that ELP flares in this work are more often associated with complex-ribbon flares. This puts into question whether the Mount Wilson classification is appropriate for judging the magnetic configuration with regard to ELP flares. The McIntosh classification for ELPs also shows generally simpler configurations to be more prone to develop ELPs, but they are also relatively uncommon.


\section{Conclusions}\label{sec:conclusions}


This work represents, to the best of our knowledge, the most extensive statistical study of EUV late-phase flares to date, with 179 ELP flares identified out of a total set of 1803 flares with a corresponding GOES class $\geq$C3.0 that occurred between May 1, 2010, and May 26, 2014. In relative terms, we see a higher probability of ELP-flare occurrence for M- and X-class flares (19\% and 16\%, respectively) than for class C~(7\%). The ratio of the late-phase and main-phase peaks ranges from 0.3 to 5.9, with 71.5\% of ELP flares exhibiting a peak ratio $>$1. This percentage of extreme ELP events is significantly higher than previously reported, due to a variety of reasons including the applied smoothing, different ELP criteria, and the usage of a pre-flare selection process. There appears to be a relationship between the late-phase delay and its duration, which is (approximately) linear. This can be explained via the longer cooling time of the loops involved in the late phase. The Mount Wilson magnetic class of ELP-harboring sunspot regions as well as their McIntosh class reveal that ELP flares, and particularly eruptive ELPs, show a slight trend toward simpler configurations as compared to the general flare population. However, previous studies reported a more complex configuration for the typical ELP flare, so the two classification schemes may not be enough to properly judge the magnetic configuration.\\
\indent Many of the findings in this work would greatly benefit from a larger sample size over a longer period of time. Especially more information on ELP flares during times of low solar activity would provide us with useful insights into the likelihood and eruptivity of ELP flares and their importance for space weather during these times. At least a coverage of one full solar cycle is desirable in order to also be able to investigate the behavior during the declining phase of a cycle. Furthermore, it would be interesting to separate ELP flares by late-phase cause and analyze the influence on the parameters described in this study (e.g., the relationship between the late-phase delay and duration).


\section{Data availability}\label{sec:data_av}


The flare list mentioned throughout this work is publicly available in the \href{https://doi.org/10.5281/zenodo.16843603}{Zenodo archive}. A description of the list entries can be found in Appendix~\ref{sec:flare_list}.

\begin{acknowledgements}
    This work has been supported by the Research Council of Norway through its Centers of Excellence scheme, project number 262622. The authors would like to thank Dr. Don Woodraska for providing access to improved EVE data. The authors acknowledge the use of data from GOES, SDO and SOHO/LASCO. GOES data are courtesy of the National Oceanic and Atmospheric Administration (NOAA). SDO is a mission for NASA’s Living With a Star program. SDO data are courtesy of NASA/SDO and the EVE and AIA science teams. SOHO is an international cooperation project between ESA and NASA. This study has benefited greatly from the open source policies of NASA and NOAA.
\end{acknowledgements}

\bibliographystyle{aa}
\bibliography{paper_bib}

\begin{appendix}

\section{Relevant emission lines}

Table~\ref{tab:lines} describes the EVE emission lines (top half) and AIA filters (bottom half) relevant for the analysis in this paper.
\begin{table}[h]
    \centering
    \caption{Selected emission lines observed by SDO/EVE and AIA with corresponding wavelength $\lambda$, primary ion species, and formation temperature $T$ (both logarithmic and absolute).}\label{tab:lines}
    \begin{tabular}{ |c|c|c|c| }
    \hline
    \multicolumn{4}{|c|}{\textbf{EVE Lines}} \\
    \hline
    $\lambda$ (nm) & Ion(s) & log$T$ & $T$ (K)\\
    \hline
    13.29 & Fe~XX & 7.0 & $10^7$ \\
    33.54 & Fe~XVI & 6.4 & $2.5 \times 10^6$ \\
    17.11 & Fe~IX & 5.8 & $6.3 \times 10^5$ \\
    30.38 & He~II & 4.8 & $6.3 \times 10^4$ \\
    \hline
    \multicolumn{4}{|c|}{\textbf{AIA Filters}} \\
    \hline
    $\lambda$ (nm) & Ion(s) & log$T$ & $T$ (K)\\
    \hline
    13.1 & Fe~VIII, Fe~XXI & 5.6, 7.0 & $4 \times 10^5$, $10^7$ \\
    33.5 & Fe~XVI & 6.4 & $2.5 \times 10^6$ \\
    17.1 & Fe~IX & 5.8 & $6.3 \times 10^5$ \\
    30.4 & He~II & 4.7 & $5 \times 10^4$ \\
    160.0 & Continuum & 3.7 & $5 \times 10^3$ \\
    \hline
    \end{tabular}
    \tablefoot{Adapted from \cite{2012SoPh..275..115W} and \mbox{\cite{2012SoPh..275...17L}}.}
\end{table}
\FloatBarrier

\section{Flare-ribbon examples}

\begin{figure}[h]
    \centering
    \includegraphics[width=0.41\textwidth]{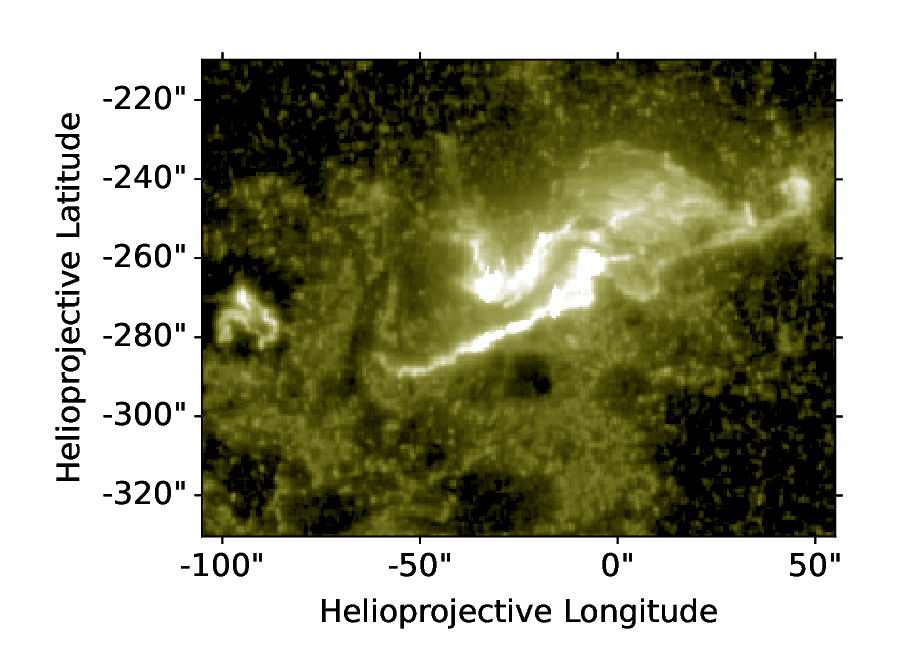}\\
    \includegraphics[width=0.41\textwidth]{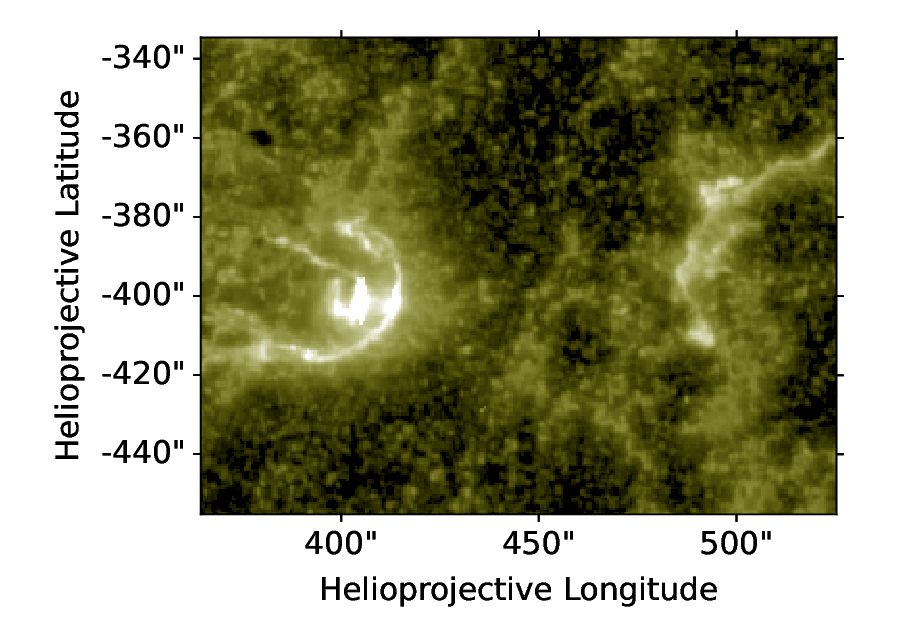}\\
    \includegraphics[width=0.41\textwidth]{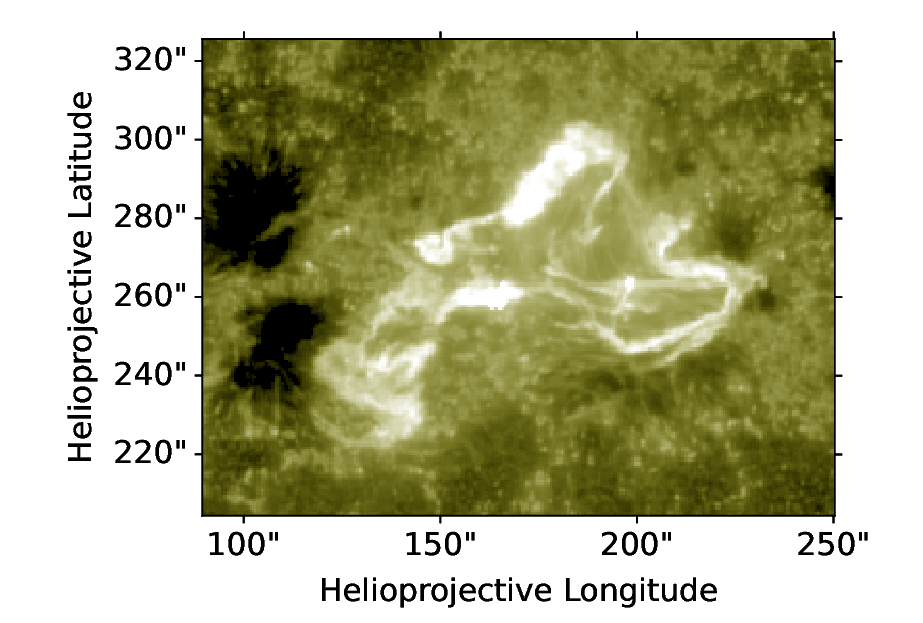}
    \caption{Examples for different flare-ribbon morphologies in AIA~160~nm images. Top panel: M6.3 two-ribbon flare on November~1,~2013 at 19:55~UT, about 2~minutes after the GOES SXR peak. This is an example of a two-ribbon morphology where ribbons can be seen to open up toward the east. Middle panel: M2.9 circular-ribbon flare on October~16,~2010. The image was taken at 19:16~UT, about 4~minutes after the peak. The circular ribbon can be seen in the left half of the image, encompassing a dot-like ribbon. To the right, a weaker remote ribbon can be observed as well. Bottom panel: X1.5 complex-ribbon flare on March~9,~2011 captured at 23:29~UT, about 6~minutes after the peak. It reveals no clear shape that could be classified as either a two-ribbon or circular-ribbon morphology.}
    \label{fig:ribbons}
\end{figure}

Figure~\ref{fig:ribbons} shows the various flare-ribbon morphologies discussed in Sect.~\ref{subsubsec:ribbons}. The top panel depicts a classical two-ribbon flare, the middle panel shows a circular-ribbon event, while the bottom panel represents a case with complex ribbons. All images were taken in the AIA~160~nm passband.
\FloatBarrier

\section{Uncertainty in the ELP duration}

\begin{figure*}[h]
    \centering
    \includegraphics[width=0.7\textwidth]{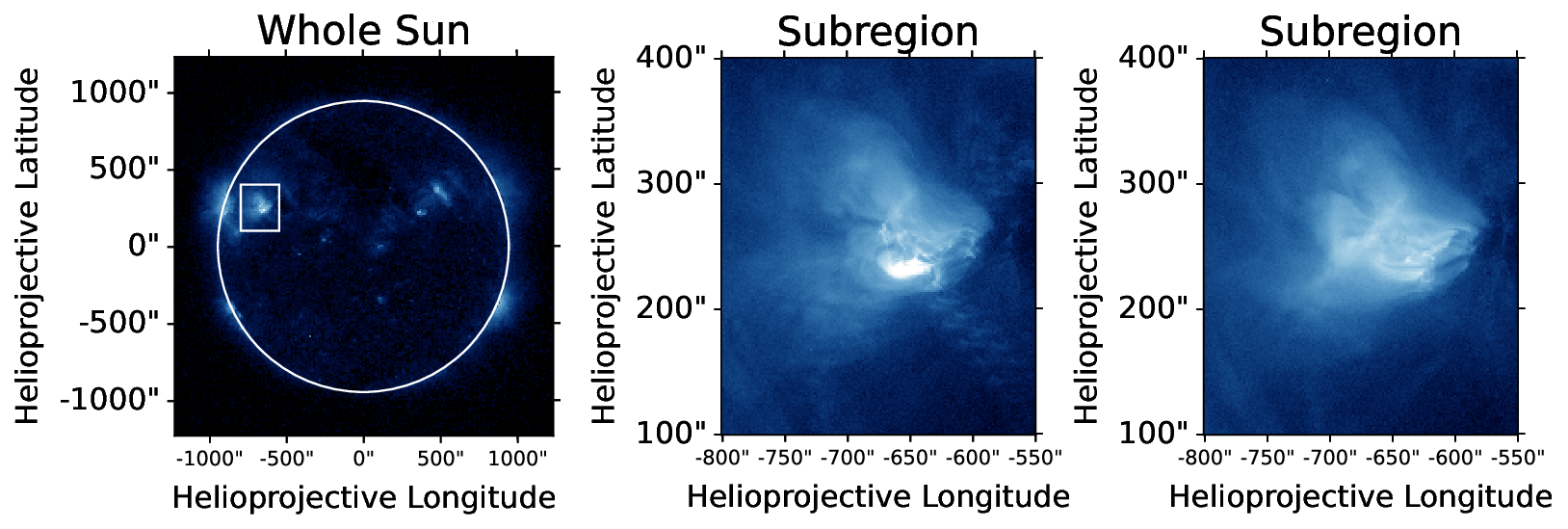}\\
    \includegraphics[width=0.7\textwidth]{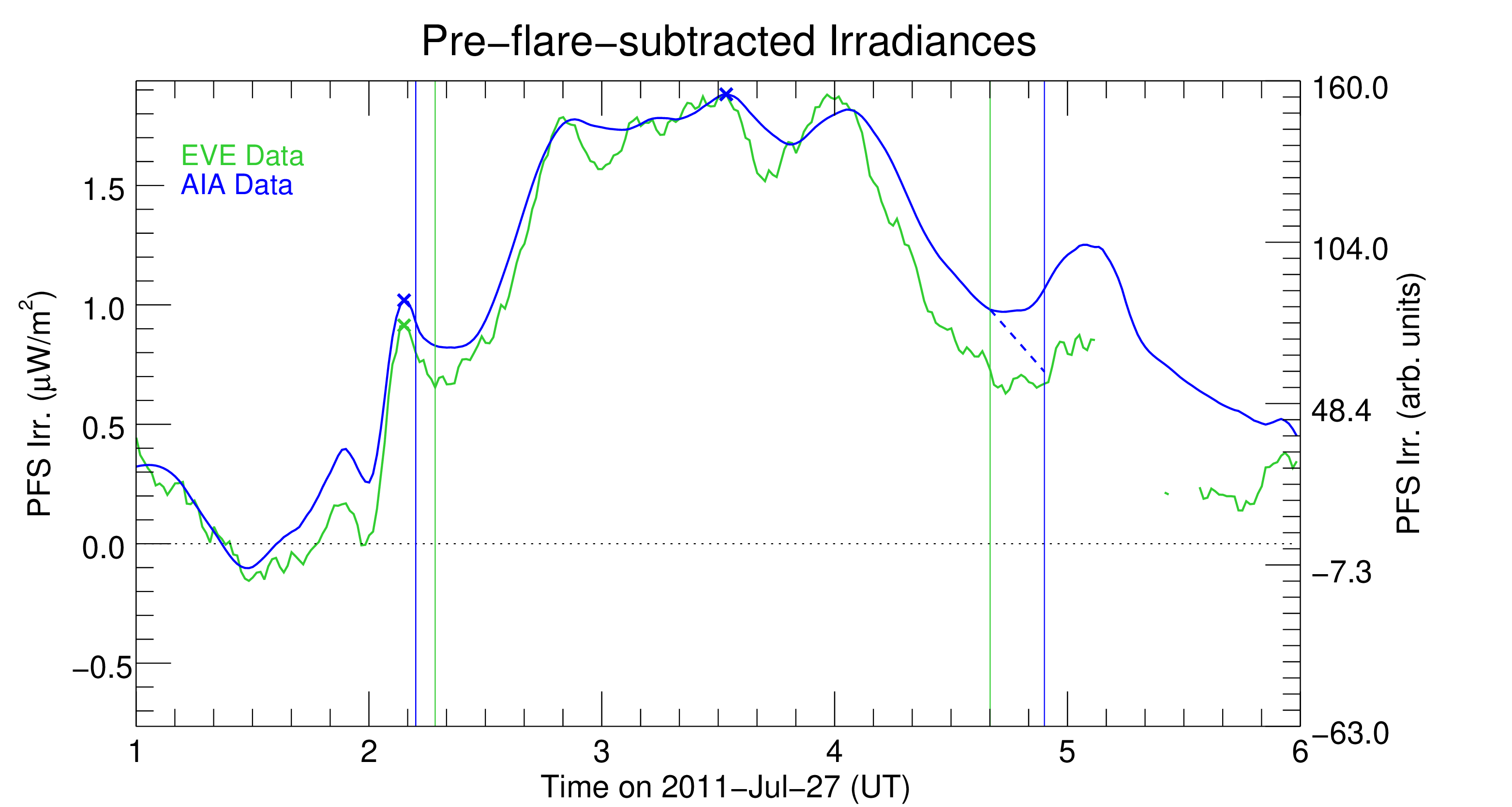}
    \caption{AIA~33.5~nm images as well as EVE and AIA light curves of a selected subregion for the C3.0~flare on July~27,~2011. The top row depicts AIA~33.5~nm images for the whole Sun (left) and the subregion (middle and right). The white box in the left image indicates the subregion, while the white circle marks the solar limb. The subregion images were taken at 02:09~UT and 03:32~UT, i.e., during the main- and late-phase peak, respectively. The bottom panel shows the EVE~33.5~nm PFS irradiance curve (green) together with the AIA~33.5~nm PFS subregion light curve (blue). The colored crosses mark the main- and late-phase peak of the respective light curve (in this case, both late-phase peaks coincide in time, so the green cross is not visible), while the colored vertical lines indicate the start and end of the ELP, as discussed in Sect.~\ref{subsubsec:duration}. The dashed blue line represents an estimate of the decrease in the emission level if no new flare was happening. The horizontal black dotted line marks the zero level.}\label{fig:eve_aia_uncertainty}
\end{figure*}

Figure~\ref{fig:eve_aia_uncertainty} depicts the determination of the duration of the ELP phase in the case of the appearance of a new flare prior to the decrease below the 40\%~threshold (see Sect.~\ref{subsec:elp_crit}) in AIA~33.5~nm emissions. The dashed blue line shows an estimate of the decrease in AIA~33.5~nm emissions if no new flare was occurring.
\FloatBarrier

\section{McIntosh class associated with ELPs}\label{sec:mcintosh_elp}

\begin{figure}
    \centering
    \includegraphics[width=0.49\textwidth]{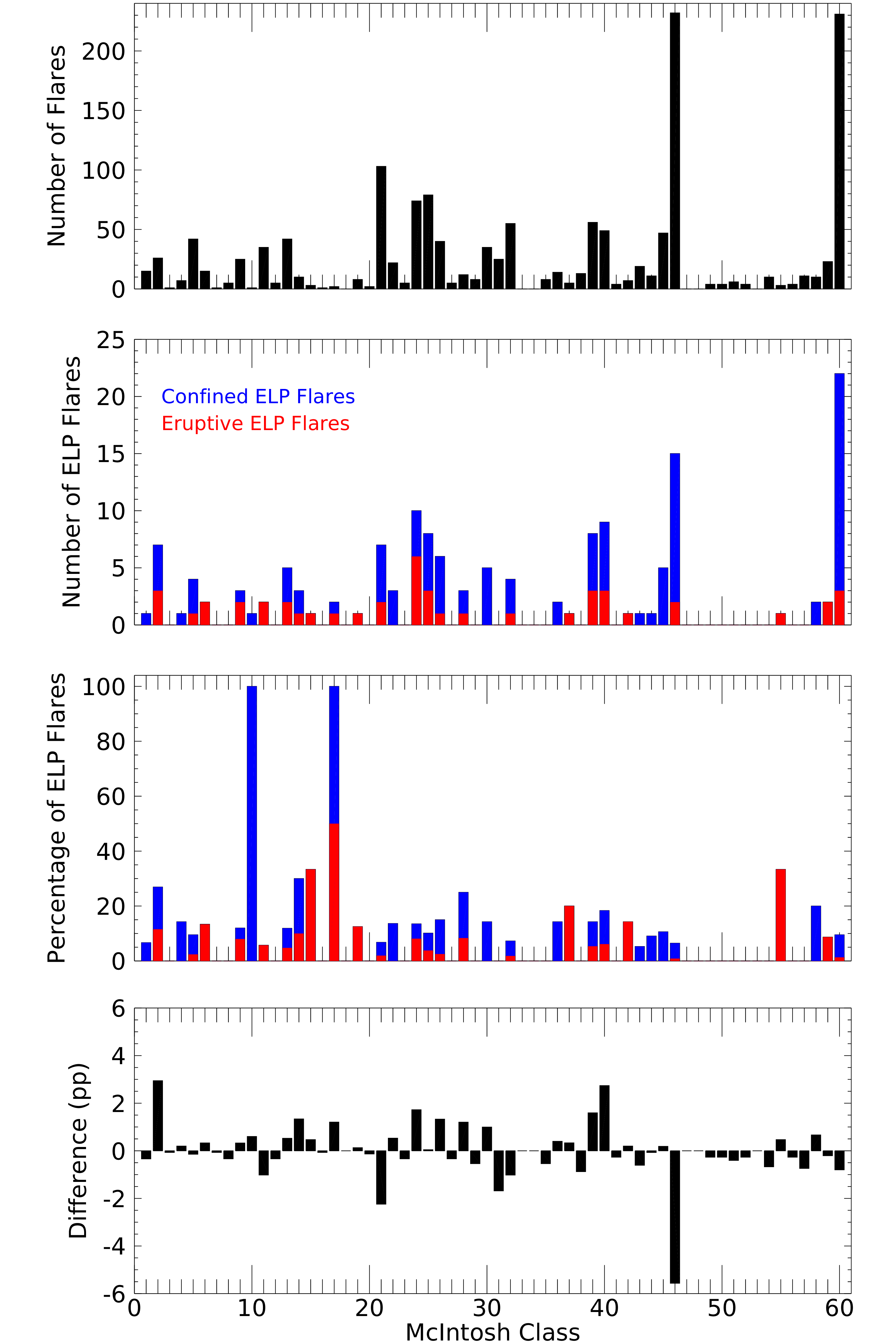}
    \caption{Number of flares $\geq$C3.0 (first panel) and number of ELP flares (second panel), percentage of flares identified as ELPs per McIntosh Class assigned to the sunspot group they appeared in (third panel), as well as difference in the distribution of McIntosh classes of sunspot regions among all flares $\geq$C3.0 vs. ELP flares. The difference is given in percentage points (pp). Positive values mean that the given class is more present in ELP flares. Eruptive ELP flares are shown in red, confined ones in blue. The numbers on the x-axis correspond to the numbers in parentheses given in Table \ref{tab:mcintosh_classes}.}
    \label{fig:mcintosh_results}
\end{figure}

Table~\ref{tab:mcintosh_classes} lists a summary of all possible McIntosh configurations. The numbers in parentheses correspond to the numbers given on the x-axis of Fig.~\ref{fig:mcintosh_results}.\\
\indent As is the case for the Mount Wilson classes, the McIntosh classification of sunspot groups also does not show a uniform distribution across all classes. In absolute terms, classes $Ekc$, $Fkc$, and $Dso$ are the most common among flares $\geq$C3.0, with 232, 231, and 103 flares created by sunspot groups belonging to these classes, respectively. The numbers for all 60~classes are shown in the top panel of Fig.~\ref{fig:mcintosh_results}. The middle two panels of Fig.~\ref{fig:mcintosh_results} depict the absolute and relative (to the overall flare population) numbers for eruptive and confined ELP flares. Again, $Fkc$ and $Ekc$ produced the largest number of ELP flares, although their roles are reversed now, with $Fkc$ being the most productive (22~ELPs) and $Ekc$ following in second place (15~ELPs).\\
\indent Relatively speaking, the difference between the distribution of McIntosh classes across all flares $\geq$ C3.0 and across ELP flares --- shown in Fig.~\ref{fig:mcintosh_results} (bottom panel) --- is less than 3 percentage points for all classes except $Ekc$. This class was responsible for 15.6\% of all flares $\geq$C3.0, but only for 10.1\% of all ELP flares. On the contrary, class $Bxo$ shows the largest increase, from 1.8 to 4.7\%. All $Da$-related classes show an increase, albeit only by about 1 percentage point on average. In total, we see a slight shift from more complex classes (right half of the graph, roughly speaking) to simpler classes (left half).
\clearpage
\begingroup
\setlength{\tabcolsep}{5.5pt}
\begin{table*}[!ht]\caption{List of the different classes within the McIntosh classification of sunspot groups.}\label{tab:mcintosh_classes}
\centering
\begin{tabular}{ccccccccc}
\toprule
\multicolumn{1}{c}{} & \multicolumn{7}{c}{\textbf{Z}} & \multicolumn{1}{c}{} \\
\cmidrule(rl){2-8}
\textbf{p} & {A} & {B} & {C} & {D} & {E} & {F} & {H} & \textbf{c}\\
\midrule
\multirow{4}{0.25cm}{x} & Axx (1) & - & - & - & - & - & - & x \\
& - & Bxo (2) & - & - & - & - & - & o \\
& - & Bxi (3) & - & - & - & - & - & i \\
& - & - & - & - & - & - & - & c \\
\midrule
\multirow{4}{0.25cm}{r} & - & - & - & - & - & - & Hrx (4) & x \\
& - & - & Cro (9) & Dro (19) & Ero (33) & Fro (47) & - & o \\
& - & - & Cri (10) & Dri (20) & Eri (34) & Fri (48) & - & i \\
& - & - & - & - & - & - & - & c \\
\midrule
\multirow{4}{0.25cm}{s} & - & - & - & - & - & - & Hsx (5) & x \\
& - & - & Cso (11) & Dso (21) & Eso (35) & Fso (49) & - & o \\
& - & - & Csi (12) & Dsi (22) & Esi (36) & Fsi (50) & - & i \\
& - & - & - & Dsc (23) & Esc (37) & Fsc (51) & - & c \\
\midrule
\multirow{4}{0.25cm}{a} & - & - & - & - & - & - & Hax (6) & x \\
& - & - & Cao (13) & Dao (24) & Eao (38) & Fao (52) & - & o \\
& - & - & Cai (14) & Dai (25) & Eai (39) & Fai (53) & - & i \\
& - & - & - & Dac (26) & Eac (40) & Fac (54) & - & c \\
\midrule
\multirow{4}{0.25cm}{h} & - & - & - & - & - & - & Hhx (7) & x \\
& - & - & Cho (15) & Dho (27) & Eho (41) & Fho (55) & - & o \\
& - & - & Chi (16) & Dhi (28) & Ehi (42) & Fhi (56) & - & i \\
& - & - & - & Dhc (29) & Ehc (43) & Fhc (57) & - & c \\
\midrule
\multirow{4}{0.25cm}{k} & - & - & - & - & - & - & Hkx (8) & x \\
& - & - & Cko (17) & Dko (30) & Eko (44) & Fko (58) & - & o \\
& - & - & Cki (18) & Dki (31) & Eki (45) & Fki (59) & - & i \\
& - & - & - & Dkc (32) & Ekc (46) & Fkc (60) & - & c \\
\midrule
\bottomrule
\end{tabular}
\tablefoot{The information is from \cite{1990SoPh..125..251M}. The numbers in parentheses signify the number of the corresponding class in Fig.~\ref{fig:mcintosh_results}.\\ Z ... Modified Zurich Class \\ p ... Penumbra of the largest sunspot \\ c ... Sunspot distribution}
\end{table*}
\endgroup
\clearpage

\section{Flare list}\label{sec:flare_list}

A list of all ELP flares between May 1, 2010 and May 26, 2014 can be found in the online \href{https://doi.org/10.5281/zenodo.16843603}{Zenodo archive}. The table describes the characteristics of all 179 ELP flares found in the aforementioned time frame. Only flares above C3.0 are taken into account. The first and second column describe the date and GOES SXR peak time ($t_{max}$) of the flare, while the third column indicates the GOES class. The fourth column describes the heliographic location on the solar disk as viewed from Earth. Column 5 expresses whether the flare was accompanied by a CME. Stars in this column indicate an uncertain flare--CME relationship. The sixth column characterizes the shape of the flare ribbons. Column 7 contains the duration of the flare ($T_{flare}$) according to GOES measurements. Columns 8 through 10 express the time delay between main-phase and late-phase maximum ($\Delta t$), the duration of the late phase ($T_{LP}$), and the ratio of the two peaks derived from the EVE~33.5~nm measurements. Uncertainties were estimated according to Sect.~\ref{subsubsec:uncertainties}. In column 9, stars mark events where no clear end of the late phase could be determined.

\end{appendix}

\end{document}